\documentclass[review]{elsarticle}

\usepackage{lineno,hyperref}
\usepackage{color}
\usepackage{subfigure}
\usepackage{mathrsfs}
\usepackage{graphicx}
\usepackage{amsmath}
\usepackage{subeqnarray}
\usepackage{cases}
\usepackage{bm}
\usepackage{amsthm}
\modulolinenumbers[5]
\newtheorem{theorem}{\textbf{\textit{Theorem}}}

\newtheorem{corollary}{\textbf{\textit{Corollary}}}
\journal{Computer Communications}

\begin{document}
\begin{frontmatter}

\title{Capacity Analysis of Public Blockchain}

\author[label1]{Xu Wang\corref{cor1}}
\ead{Xu.Wang-1@uts.edu.au}
\cortext[cor1]{Corresponding author.}
\author[label2]{Wei Ni}
\ead{Wei.Ni@data61.csiro.au}
\author[label3]{Xuan Zha}
\ead{zhaxuan@caict.ac.cn}
\author[label1]{Guangsheng Yu}
\ead{Guangsheng.Yu@uts.edu.au}
\author[label1]{Ren Ping Liu}
\ead{RenPing.Liu@uts.edu.au}
\author[label4]{Nektarios Georgalas}
\ead{nektarios.georgalas@bt.com}
\author[label4]{Andrew Reeves}
\ead{andrew.reeves@bt.com}
\address[label1]{Global Big Data Technologies Centre, University of Technology Sydney, Australia}
\address[label2]{Data61, CSIRO, Australia}
\address[label3]{China Academy of Information and Communications Technology, Beijing, China}
\address[label4]{Applied Research, British Telecom, Martlesham, UK}

\begin{abstract}
As distributed ledgers, blockchains run consensus protocols which trade capacity for consistency, especially in non-ideal networks with incomplete connectivity and erroneous links.
Existing studies on the tradeoff between capacity and consistency are only qualitative or rely on specific assumptions.
This paper presents discrete-time Markov chain models to quantify the capacity of Proof-of-Work based public blockchains in non-ideal networks. The comprehensive model is collapsed to be ergodic under the \emph{eventual consistency} of blockchains, achieving tractability and efficient evaluations of blockchain capacity. A closed-form expression for the capacity is derived in the case of two miners.
Another important aspect is that we extend the ergodic model to analyze the capacity under \emph{strong consistency}, evaluating the robustness of blockchains against double-spending attacks.
Validated by simulations, the proposed models are accurate and reveal the effect of link quality and the distribution of mining rates on blockchain capacity and the ratio of stale blocks.
\end{abstract}


\end{frontmatter}


\section{Introduction}
\label{sec_introduction}
Blockchain provides a distributed tamper-resistant ledger system, which chains blocks of transactions by cryptographical means, stores and updates the chain at distributed locations~\cite{WANG201910}.
Blockchain is the enabling technique underlying the proliferating cryptocurrencies, such as Bitcoin~\cite{Nakamoto:2008ti}, and has been developed to provide tamper-resistant services~\cite{moro2020distributed}.
Blockchain miners verify transactions, which are created and signed by blockchain users and then propagated across the peer-to-peer blockchain network, and assemble the transactions into blocks in a decentralized manner.
The blocks are chained together with their cryptographic hash values and kept at every individual miner as a local record.

Consensus protocols are run by decentralized miners to mitigate forks resulting from inconsistent chains at different locations~\cite{kogias2016enhancing}.
The most popular consensus protocol in public blockchains is Proof-of-Work (PoW), such as in Bitcoin and Ethereum~\cite{wood2014ethereum,10.1007/978-3-662-46803-6_10}.
A miner first assembles the hash value of the latest block in its local chain, transactions and block metadata into a block.
The miner needs to continuously change the nonce in the block until the block hash value is less than a global target, and then broadcasts the block with the valid nonce across the network~\cite{Nakamoto:2008ti}.
The blocks mined by different miners are different because of inconsistent views on local chains and transactions, but require the same level of effort to meet the same global target.
Apart from PoW, other consensus protocols in public blockchains include Proof-of-Stake (PoS)~\cite{king2012ppcoin} and Proof-of-Activity (PoA)~\cite{Bentov:2014:PAE:2695533.2695545}.
Different from public blockchains, private blockchains operate in well-controlled networks and adopt negotiation-based protocols, e.g., Practical Byzantine Fault Tolerance (PBFT)~\cite{Castro:1999te}, to achieve consensus among a relatively small number of miners.

In blockchains, capacity defines the highest rate at which transactions are admitted into the globally endorsed blockchain. The blockchain capacity is limited by the consistency requirement, such as \emph{strong consistency} and \emph{eventual consistency}, as distributed ledgers~\cite{vogels2009eventually,6688704}. Specifically, the block mining process needs to be slowed down, to mitigate inconsistency across the blockchain network. This, unfortunately, sacrifices capacity. Such tradeoff between capacity and consistency can be particularly severe in public networks where PoW based public blockchains are typically adopted. The reason is that the non-ideal connections of public networks, such as incomplete connectivity, multi-hop delays, and non-deterministic loss per hop, would make it take longer to deliver blocks across the networks, or even prevent successful deliveries~\cite{7931680}. The capacity would be penalized to preserve consistency. 
Moreover, the blocks may not be delivered in time due to the non-ideal network connections, resulting in inconsistency across the blockchain network~\cite{6688704}.
The globally endorsed blockchain solves the inconsistency by keeping the blocks in the global longest chain and discarding any other blocks.
This strategy would further reduce the blockchain capacity~\cite{7423672}.
Take Bitcoin as an example, a block is no larger than one Megabyte and generated every ten minutes on average to preserve consistency. As a consequence, the capacity of Bitcoin is, on average, six blocks per hour~\cite{6688704}.

The blockchain capacity and blockchain extension process have been quantitatively analyzed under assumptions that simplify the complicated interactions among miners.
Some works ignore the inconsistency to decouple the block mining process across miners. For example, miners were assumed to stick on their local chains and rarely update chains with counterparts~\cite{Nakamoto:2008ti,piriou2018simulation,YANG2020101956} or always have consistent views on the public blockchain~\cite{YU2020101934}. 
The inconsistency was evaluated by the ratio of discarded blocks (i.e., the stale block rate) that can be affected by many factors, such as block intervals, network delay, information propagation mechanisms and network configuration~\cite{Gervais:2016tu}.
Some studies employed stale block rates from simulations or empirical stale block rates to analyze blockchain security and performance~\cite{Gervais:2016tu,YANG2020101956}. 
The stale block rate was estimated under a bounded block propagation delay among miners and the assumption that a fork occurs if new blocks are mined before preceding blocks have propagated to all miners~\cite{sompolinsky2015secure,8946275,10.1145/3412341}. This assumption would sacrifice accuracy because the new blocks can be mined by the miners that have learned the latest blockchain.

This paper presents new Markov chain models that can accurately quantify the capacity and consistency of PoW-based public blockchains and capture the impact of non-ideal links on the blockchain capacity. We start by modeling under the \emph{eventual consistency} requirement of public blockchains. A comprehensive Discrete-Time Markov Chain (DTMC) model is established, where the lengths of chains at miners are captured in Markov statuses, and then compressed to be an Ergodic DTMC (EDTMC) for numerical evaluations of generic network settings. 
A closed-form expression is derived for the capacity of a two-miner blockchain system with unreliable links.
We further generalize the EDTMC model to evaluate the capacity of the blockchain system in the state of \emph{strong consistency} which is important for instantly verifying transactions and preventing double-spending attacks.

The contributions of this paper are summarized as follows.
\begin{itemize}
\item{We propose a DTMC model to characterize blockchains in non-ideal networks, which is collapsed to be an EDTMC model under the eventual consistency to improve the computational feasibility;}
\item{A closed-form expression is derived for the blockchain growth rate and the stale block rate of two-miner blockchain systems in non-ideal networks;}
\item{We extend the EDTMC model under the strong consistency to trace the block miners in addition to the blockchain growth rate. A closed-form expression is derived in the ideal case of two miners which are connected by error-free and reliable links.}
\end{itemize}

Validated by simulations, the proposed models are able to quantify the impacts of mining rates of the miners and link conditions on the blockchain capacity.
We show that the blockchain capacity is strongly affected by the distribution of the mining rates and lower-bounded by the case with even mining rates.

The rest of this article is organized as follows.
Section~\ref{sec_relatedwork} studies related works, followed by the system model and definition in Section~\ref{sec_sysmodel}.
Section~\ref{sec_genmodel} presents the DTMC/EDTMC models for the blockchain capacity analysis under the eventual consistency, followed by the closed-form expression for the capacity of two miners connected by non-ideal links in Section~\ref{sec_captwominers}.
In Section~\ref{sec_stongcon}, the strong consistency of a public blockchain is evaluated.
The proposed models are numerically validated in Section \ref{sec_simwithinlan}, followed by conclusions in Section~\ref{sec_con}.

\section{Related Work}
\label{sec_relatedwork}
Most existing theories on the capacity and consistency of a network, such as the well-known Brewer's theorem~\cite{Gilbert:2002il} and PACELC theorem~\cite{6127847}, are qualitative and cannot be readily applied to quantitatively evaluate the capacity of public blockchains. 
Some models have been designed to quantify the consistency and capacity of PoW-based public blockchains~\cite{Nakamoto:2008ti,sompolinsky2015secure,8946275,10.1145/3412341}. These models rely on specific assumptions that simplify~\cite{sompolinsky2015secure,8946275,10.1145/3412341} or decouple~\cite{Nakamoto:2008ti} the blockchain extension process across miners and thus compromise model accuracy.

Brewer's theorem, also widely known as the Consistency, Availability and Partition tolerance (CAP) theorem, was established in 2002~\cite{Gilbert:2002il}. The theorem extracts three vital guarantees for general distributed systems, namely, consistency, availability and partition-tolerance; and dictates qualitatively that it is impossible for a distributed system to simultaneously provide all the three guarantees in the meantime.
The CAP theorem is recently extended to be the PACELC theorem~\cite{6127847} pointing out the overlook of the consistency-latency trade-off in the CAP theorem and revealing the existence of the trade-off between latency and consistency even in the absence of partitioning.
Although insightful, the theorems do not provide a quantitative understanding of the trade-offs and practical applications~\cite{rahman2017characterizing}.
The consistency referred in the theorems is single-copy consistency~\cite{6133253} or linearizability~\cite{Gilbert:2002il}, which is a type of \emph{strong consistency}~\cite{kleppmann2015critique}. A blockchain is considered to provide \emph{strong consistency} and serializability by delaying confirmation of the latest blocks in the blockchain. For example, Bitcoin is strongly consistent with high probability because of the six-confirmation rule~\cite{sirer2016bitcoin}.
The physical characteristics of networks and links, such as latency, finite link bandwidth and erroneous link, can be parametrized as a threshold; which would lead to partitioning a system into isolated subsystems, if violated~\cite{6133253}. However, this threshold has an implicit assumption of separable networks, while many networks are inseparable~\cite{wang2016virus}.

By exploiting distributed consensus protocols, public blockchains have to trade off between capacity and consistency, as dictated in the CAP theorem. Blockchain simulation platforms, such as BlockSim~\cite{10.1145/3308897.3308956} and SimBlock~\cite{8845253}, have been developed to investigate the public blockchain growth from experimental approaches. The time-consuming PoW mining process is simulated with probability functions. The blockchain parameters, such as the number of nodes, block size and block interval, and network parameters, e.g., propagation delay and bandwidth, can be configured to evaluate the impact on the blockchain extension.

Blockchain capacity and consistency have been studied by modeling the block mining and chain extension process.
In Nakamoto's paper~\cite{Nakamoto:2008ti}, the blockchain extension at a benign miner and an attacker was formulated with a Binomial Random Walk model to deduce the attacker's winning probability. The model assumed that the miner and the attacker stuck on their local chains and ignored the interaction between the miner and the attacker. However, the blockchain extension among multiple benign miners is more complicated. Miners may have inconsistent views of the blockchain due to non-ideal connections and continuously switch among the chains (according to the chain selection rule).
The impact of inconsistency across miners can be evaluated by the ratio of discarded blocks (i.e., the stale block rate). In~\cite{Gervais:2016tu}, the stale block rates from observation and simulator were fed into Markov Decision Process (MDP) models to analyze the security and performance against selfish mining and double-spending attacks.
Instead of carrying out time-consuming simulations of stale block rates~\cite{Gervais:2016tu}, there have been attempts to quantitatively analyze the stale block rate~\cite{8946275,10.1145/3412341,sompolinsky2015secure}.
By assuming that the blocks are stale if they are mined before the preceding blocks propagate to all miners within a bounded block propagation delay, the stale block rate was presented in~\cite{8946275}. The bounded block propagation delay can be quantified by formulating the Gossip protocol, e.g., in~\cite{10.1145/3412341}. 
This assumption can overestimate the stale block rate and thus underestimate the blockchain capacity because the blocks can be mined by the miners who have learned the latest blockchain and are nonstale.
By developing the theory on Markovian systems with a constant delay between two miners, a closed-form expression was developed for the blockchain capacity~\cite{sompolinsky2015secure}. However, the model and result have some limitations, such as only two miners, unequal mining rates and a bidirectional and constant delay.

The Markov theory has been developed to model other dynamic processes in blockchain where blocks are assumed to be generated at a fixed rate as an inherent blockchain parameter. 
In~\cite{li2019markov}, the transaction arrivals and mining process were analyzed with a Markov process of GI/M/1 type. The number of unconfirmed transactions, i.e., the number of transactions in the transaction pool, and the transaction confirmation time were calculated with the Markov model. 
In~\cite{motlagh2020modeling}, a Markov chain was developed to capture the block synchronization process of churning nodes that may leave the blockchain network and rejoin after some time. The synchronization time and block distribution time were derived from the Markov chain.

In our recent work~\cite{WANG2019100109}, a DTMC model was proposed to capture the blockchain growth under the eventual consistency for the developed blockchain-IoT testbed. The infinite Markov model of the DTMC was approximated by a finite Markov model to evaluate the steady states and the blockchain capacity under the eventual consistency. However, the model only considered the eventual consistency and thus failed to capture the strong consistency which is critical to analyzing the double-spending attacks in blockchains.
Meanwhile, the DTMC model in~\cite{WANG2019100109} only gave numerical upper bounds rather than accurate closed-form results.

\section{System Model and Definition}
\label{sec_sysmodel}
We consider public blockchains where blocks are mined in parallel based on consensus problems at different miners. 
Miners can solve the consensus problems simultaneously and mine more than one blocks that have the same index. These blocks conflict with each other and can cause forks, although they may contain the same set of transactions.
Only one of the conflicting blocks can be accepted in a blockchain which acts as a distributed ledger and pursues \emph{eventual consistency}. The other blocks are discarded.

The blockchain capacity defines the highest rate at which transactions are admitted into the globally endorsed blockchain.
The blockchain capacity, denoted by $R$, can be given by
\begin{equation}
R=\delta\times\zeta,
\end{equation}
where $\delta$ is the extension rate of the globally endorsed chain (i.e., the number of blocks per unit time appended to the longest of the chains maintained at distributed locations).
The maximum number of transactions per block is denoted by $\zeta$ and can be calculated based on the block size limit and the minimum transaction size. Without loss of generality, we assume $\zeta=1$. The blockchain capacity can be assessed as the extension rate of the globally endorsed blockchain.

We consider a public blockchain system with $n$ miners operating on a slotted basis (e.g., running the network timing protocol (NTP)~\cite{103043}).
The timeslot is short enough that a miner can mine no more than a block per timeslot.
This is due to the fact that the block mining of a miner is a sequential process because mining a block needs the hash value of the preceding block and, therefore, can only be carried out after mining the preceding block.
Multiple miners can mine different blocks in parallel in the same timeslot on top of their local chains.
We use $\mathbf c=[c_{1},c_{2},\cdots,c_{n}]$ to denote the mining rates of $n$ miners, where $c_{i}$ gives the probability that the $i$-th miner produces a block (and appends the block to its own local chain) in any timeslot.

\begin{figure}[!h]
\centering
\includegraphics[width=2in]{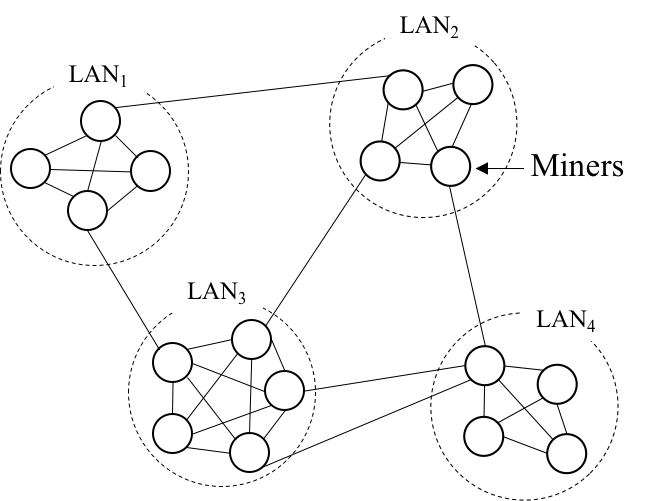}
\caption{An example of a public blockchain network, where miners can be connected by non-ideal links with error and delay. The network can be partitioned into multiple segments, due to the non-ideal links.}
\label{fig_overallTopo}
\end{figure}

In an ideal scenario where the miners are connected by error-free links, the miners could synchronize with their neighbors in terms of their knowledge on the longest blockchains in every timeslot. For example, the miners within a local area network (LAN) can have such error-free, ideal links in-between, as shown in Fig.~\ref{fig_overallTopo}.
The blockchain capacity of the $n$ miners connected by ideal links, denoted by $\bar R$, can be calculated with their mining rates,
\begin{equation}
\label{equ_etg}
\Bar R=1-\prod_{i=1}^n(1-c_i).
\end{equation}
This is also the probability that at least one block is mined by different miners in a timeslot. In the ideal scenario, all miners have learned the longest chains and independently try to extend the chains by a block. In a timeslot, more than a block can be mined by different miners. These blocks are at the same chain height, and only one of the blocks is eventually admitted in the globally endorsed blockchain according to the longest chain rule. The other blocks that are not admitted are discarded. As a result, the globally endorsed blockchain can grow by a block with probability $\bar R$.

In non-ideal networks, the connections among miners can be unreliable, preventing the miners from getting instant views on the longest chain in time.
Let $a_{ij}$ and $d_{ij}$ be the probability that the $i$-th miner can successfully pass its knowledge of the longest chain to the $j$-th miner in a timeslot and the delay of the link, respectively. They describe the non-ideal links between the miners, as given by~\cite{durrett2010probability}
\begin{equation}
\label{equ_delayalpha}
d_{ij}=\sum_{k=1}^\infty a_{ij} k(1-a_{ij})^{k-1}=\frac{1}{a_{ij}},
\setcounter{equation}{3}
\end{equation}
where $k$ indicates failed transmissions in the preceding $k-1$ timeslots and a successful transmission in the $k$-th timeslot. In other words, a message needs an average $d_{ij}=\frac{1}{a_{ij}}$ (re)transmissions to be successfully delivered through the link with the transmission success rate of $a_{ij}$.

One example is given in Fig.~\ref{fig_overallTopo} that miners locate in separated LANs that are connected with non-ideal links. Miners cannot always get instant views of chains from miners in other LANs. The miners would mine blocks based on their local chains. When the miners learn chains from other miners, they would compare their local chains and received chains and then keep the longest one following the longest chain rule. The interaction among miners and the extension of the globally endorsed blockchain can be complicated due to decentralized running of the longest chain rule and the switch between chains and non-ideal links, i.e.,~$a_{i,j}$.

Stale blocks refer to the blocks that are temporarily appended to local chains at some miners (not the global longest chain) and discarded later when the local chains are replaced by the global longest chain. Stale blocks are inevitable in decentralized blockchain systems and can greatly reduce the security performance of the blockchain~\cite{Gervais:2016tu}.
The ratio of stale blocks, denoted by $O_r$, can be readily inferred from the blockchain capacity $R$, as given by
\begin{equation}
\label{equ_ratioStale}
O_{r}=\frac{\sum_{i=  1}^{n} c_{i}-R}{\sum_{i=1}^{n} c_{i}}.
\end{equation}

Notations used in this paper are collated in Table~\ref{tab_notation}, and model assumptions are collated as follows,
\begin{itemize}
    \item The blockchain follows the longest chain rule, i.e., the longest chain of all miners is selected as the globally endorsed blockchain;
    \item The system is discrete-time, where each timeslot consists of a synchronization phase and a mining phase;
    \item In any timeslot, a miner can mine no more than one block, and different miners can independently mine blocks; 
    \item The links among the miners can be non-ideal and described with transmission success probabilities;
    \item Any blockchain transaction can be instantly processed.
\end{itemize}

\begin{table}[!htb]
\caption{Notations and Definitions}
\label{tab_notation}
\begin{tabular}{l|l}
\hline
\textbf{Notation} & \textbf{Definition}                                                                                             \\ \hline
$R$               & Blockchain capacity                                                                                             \\
$\bar R$          & The blockchain capacity of $n$ miners connected by ideal links                                                  \\
$O_{r}$           & The ratio of stale blocks                                                                                       \\
$\delta$          & The growth rate of the blockchain in each timeslot                             \\
$\zeta$           & The number of transactions per block          \\
$n$ & The number of miners\\
$c_i$             & The mining rate of the $i$-th miner                                                                             \\
$a_{i,j}$         & The transmission probability from the $i$-th to the $j$-th miner\\
$\mathcal{S}$     & The synchronization phase in a timeslot                                                                         \\
$\mathcal{M}$     & The mining phase in a timeslot                                                                                  \\
$\bm b$           & The status of the DTMC \\
$b_i$ & The number of blocks in the local chain at the $i$-th miner                    \\
$\bm r$           & The status of the EDTMC\\
$r_i$ &The relative length of the local chain at the $i$-th miner                                     \\
$\pi_{\bm r}$     & The steady-state probability of the status $\bm r$                                                              \\
$\mathbf V$       & The status of the FDTMC \\
$v_{i,j}$ & The miner of the $j$-th block in the $i$-th miner's local chain \\
$\tau_{i,j}$ & The steady-state probability of the status $[\mathbf 1_{i},2\times \mathbf 1_{j}]$\\
$\gamma_i$        & The ratio of admitted blocks mined by the $i$-th miner
\\ \hline
\end{tabular}
\end{table}

\section{Capacity Analysis under Eventual Consistency}
\label{sec_genmodel}
In this section, we start with a Discrete-Time Markov Chain (DTMC) which captures the PoW blockchain dynamics in non-ideal networks with $n$ miners. From the perspective of the \emph{eventual consistency}, we focus on the blockchain capacity and the stale block rate and ignore the specific miners of blocks in the globally endorsed blockchain. 

\begin{figure}[!h]
\centering
\includegraphics[width=3in]{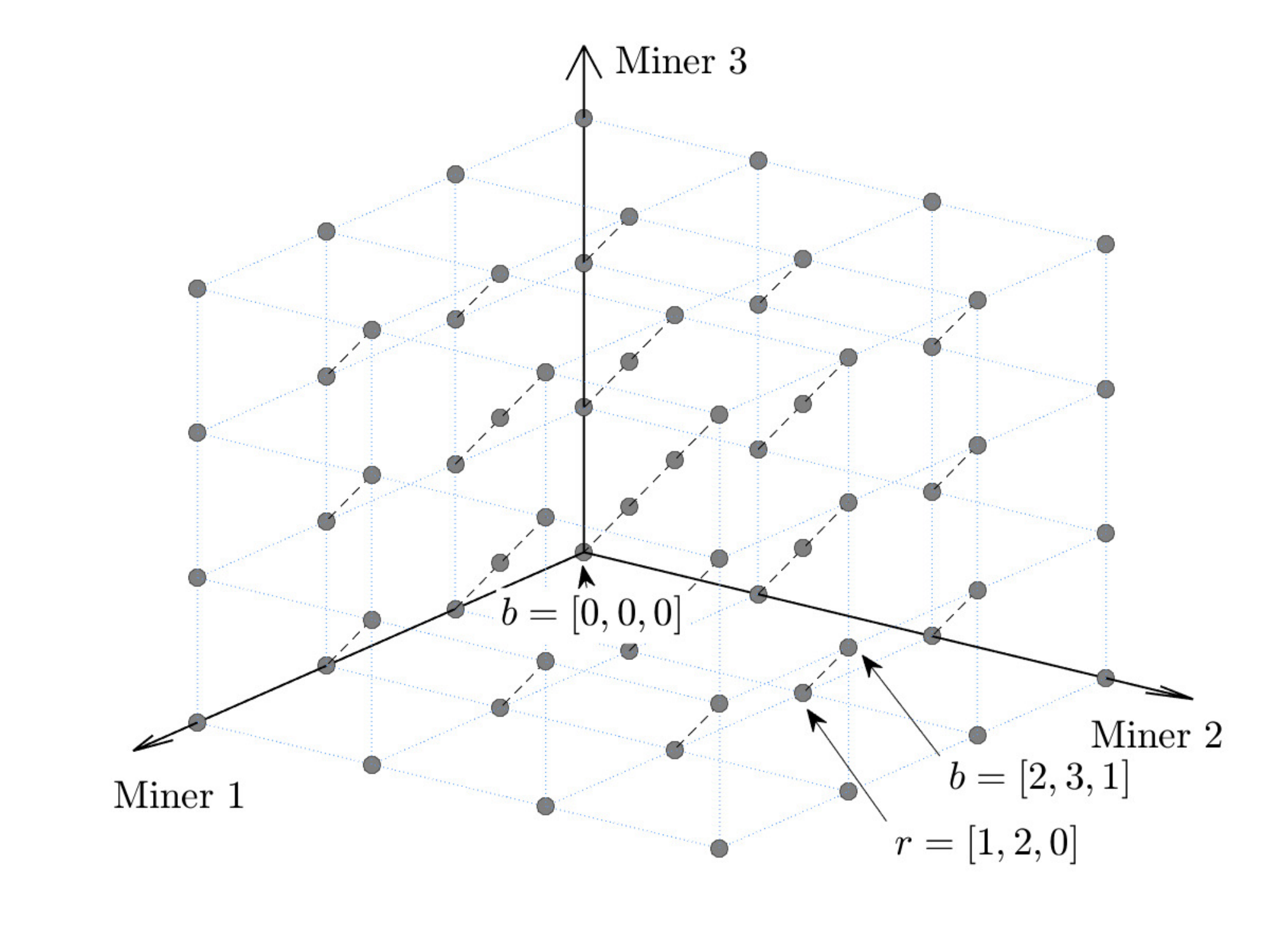}
\caption{The statuses of the DTMC/EDTMC model. For example, $\bm b=[2,3,1]$ represents that the three miners have 2, 3, and 1 blocks, which can be collapsed to be $\bm r=[1,2,0]$, indicating that the first and second miners have 1 and 2 more blocks than the third miner.}
\label{fig_threeNodesExample1}
\end{figure}

In any timeslot, the $i$-th miner can successfully pass its local chain to the $j$-th miner with probability $a_{ij}$ as defined in \eqref{equ_delayalpha}.
The status of the chains can be represented with $\bm b=[b_1, b_2, \cdots, b_n]$, where $b_i$ is the number of blocks in the local chain at $i$-th miner. The status of the blockchain network with three miners is illustrated in Fig. \ref{fig_threeNodesExample1}.

In a timeslot, the chain extension includes a \emph{Synchronization} ($\mathcal{S}$) phase, where miners exchange local chains with connected miners and then only keep the longest one and drop the others. The chain extension also includes a \emph{Mining} ($\mathcal{M}$) phase, where the miners independently solve the global mining problems to mine blocks and append the mined blocks to the end of their local chains.
We assume that the timeslot is small enough that a miner can mine at most one block. The transition probabilities of the DTMC can be formulated by
\begin{equation}\label{eq: trans probability full}
\Pr\{\bm b'|\bm b\} = \sum_{\bar {\bm b}} \Pr\{\bar{\bm b}\big|\bm b,\mathcal{S}\} \Pr\{\bm b'\big|\bar {\bm b},\mathcal{M}\}.
\end{equation}
Here, $\bar{\bm b}$ is an intermediate status after the $\mathcal{S}$ phase. $\Pr\{\bar{\bm b}\big|\bm b,\mathcal{S}\}$ gives the transition probability from status $\bm b$ to $\bar{\bm b}$ via the $\mathcal{S}$ phase, and $\Pr\{\bm b'\big|\bar{\bm b},\mathcal{M}\}$ gives the transition probability from $\bar {\bm b}$ to $\bm b'$ across the $\mathcal{M}$ phase.

Since every miner independently updates its chain based on its knowledge of its neighbors, $\Pr\{\bm {\bar b}\big|\bm b,\mathcal{S}\}$ can be given by
\begin{equation}\label{eq: trans probability sync}
\Pr\{\bar{\bm b}\big|\bm b,\mathcal{S}\}=\prod_{i=1}^{n}\Pr\{\bar b_{i}|b_{i},\mathcal{S}\}.
\end{equation}
The transition probability $\Pr\{\bar b_{i}|b_{i},\mathcal{S}\}$ is the probability that the $i$-th miner replaces its local chain with the length of $b_{i}$ by a chain with the length of $\bar b_{i}$ from its neighbors, and can be given by
\begin{subequations}
\label{eq_ps}
\begin{numcases}{\Pr\{\bar b_{i}|b_{i},\mathcal{S}\}=}
\left(1-\underset{\substack{j=1\\b_{j}=\bar b_{i}}}{\overset{n}{\prod}}(1-a_{ji})\right)\underset{\substack{j=1\\b_{j}>\bar b_{i}}}{\overset{n}{\prod}}(1-a_{ji}),\notag\\
\text{ if there exists } k\neq i\text{ and }b_{k}=\bar b_{i}>b_{i};&\label{eq_ps1}\\
\underset{\substack{j=1\\b_{j}>\bar b_{i}}}{\overset{n}{\prod}}(1-a_{ji}),\notag\\
\text{ if there exists } k\neq i\text{ and }b_{k}>\bar b_{i}=b_{i};&\label{eq_ps2}\\
1,\qquad\qquad\qquad\quad\text{if }\bar b_{i}=b_{i}>b_{k},\ \forall k;&\label{eq_ps4}\\
0,\qquad\qquad\qquad\qquad\qquad\ \text{otherwise}.&\label{eq_ps3}
\end{numcases}
\end{subequations}
Here, \eqref{eq_ps1} captures the case that the $i$-th miner receives a chain with $\bar b_i$ blocks from its neighbors and uses one of them to replace the shorter local chain.
\eqref{eq_ps2} indicates that the miner keeps its $b_{i}$-block long local chain if it does not receive any chain longer than $b_{i}$.
\eqref{eq_ps4} is because the miners having the longest local chain keep their chains.

In the mining phase, miners independently mine blocks and extend their local chains. Thus, we have  
\begin{equation}
\Pr\{\bm b'\big|\bm {\bar b},\mathcal{M}\}=\prod_{i=1}^{n}\Pr\{ b_{i}'\big|\bar b_{i},\mathcal{M}\}.
\end{equation}
Here, the number of blocks in $i$-th miner's chain extends from $\bar b_i$ to $b'_i$ with probability $\Pr\{b_{i}'\big| \bar b_{i},\mathcal{M}\}$, which is given by
\begin{equation}\label{equ_pms}\begin{split}
\Pr\{b_{i}'\big| \bar b_{i},\mathcal{M}\}=\begin{cases}
c_{i},&\text{if }b'_{i}=\bar b_{i}+1;\\
(1-c_{i}),&\text{if }b'_{i}=\bar b_{i};\\
0,&\text{otherwise}.
\end{cases}\end{split}\end{equation}
We can also conclude that 
\begin{equation}\label{eq. proof 1}
\Pr\{\bm b'+k\times\mathbf 1_{n}|\bm b+k\times\mathbf 1_{n}\}=\Pr\{\bm b'|\bm b\},\  \forall k\in Z^+,
\end{equation}
where $Z^+$ and $\mathbf 1_{n}$ denote positive integers and the $n$-dimensional all-one vector, respectively.

After building up the DTMC, we collapse the diagonal statuses for an ergodic DTMC (EDTMC) applying the \emph{eventual consistency} feature of blockchain.
The statuses of the EDTMC, denoted by $\bm r$, represent the relative chain lengths rather than the actual lengths of local chains at different miners. This is achieved by letting $\bm r=\bm b -\min(\bm b)\times\mathbf 1_{n}$, where $\min(\bm b)$ captures the shortest length of the local chains (i.e., the minimum element of $\bm b$). For example, $\bm b=[2,3,1]$ can be collapsed to be $\bm r=[1,2,0]$, as shown in Fig.~\ref{fig_threeNodesExample1}.

The EDTMC preserves the Markov property because only the relative lengths of the chains are used to evaluate the transition probabilities in \eqref{eq_ps} and \eqref{equ_pms}. The transition probabilities of the EDTMC can be given by 
\begin{equation}
\label{equ_uandp}
\begin{split}
&\Pr\{\bm r' |\bm r,\mathcal{E}\}=\sum_{k=0}^{\infty}\Pr\{\bm r'+k\times\mathbf 1_{n}| \bm r\}.
\end{split}
\end{equation}
Here, $\mathcal E$ indicates that this is the transition probability for the EDTMC model, and $\Pr\{\bm r'+k\times\mathbf 1_{n}| \bm r\}$ can be obtained with \eqref{eq: trans probability full}.

The proposed EDTMC model with infinite statuses has unique steady-state probabilities because the blockchain network can reach the status of $\bm r_0=[0,0,\cdots,0]$ when all the local chains at different miners have the identical number of blocks~\cite[Theorem 16]{konstantopoulos2009markov}. With the steady states of the EDTMC model, the blockchain capacity can be given by 
\begin{equation}
\label{equ_rpiw}
R=\sum_{\bm r}(\pi_{\bm r}\times \omega_{\bm r}).
\end{equation}
Here, $\pi_{\bm r}$ is the steady-state probability of the status $\bm r$. $\omega_{\bm r}$ is the expected blockchain capacity of status $\bm r$ and can be given by 
\begin{equation}
\label{sec_ww}
\omega_{\bm r}=\sum_{\bm b}\left(\Pr\{\bm b\big|\bm r,\mathcal{S}\}\left(1-\underset{\substack{i=1\\b_{i}=\max(\bm b)}}{\overset{n}{\prod}}(1-c_{i})\right)\right).
\end{equation}
This captures the case that at least one miner with the longest local chains after the $\mathcal{S}$ phase mines a block and therefore extends the global longest chain.

The EDTMC with infinite statuses can be asymptotically approximated with a Markov model with finite statuses to improve the computational feasibility. We assume that the longest chain can only exceed up to $k$ blocks at most. In practice, Bitcoin recommends $k=6$ that blocks earlier than six blocks have been eventually endorsed by all the miners~\cite{Nakamoto:2008ti}. The approximation Markov model takes the statuses and transition probabilities from the EDTMC when the maximum relative length of chains is less than $m$, i.e.,  $\max(\bm r)<k$. In the case of $\max(\bm r)=k$, the status is redirected to a status $\bm r'$ with $\max(\bm r')<k$. The status $\bm r'$ is given by
\begin{equation}
\label{equ_appro}
r'_{i}=
\begin{cases}
r_{i}-1,&r_{i}>0;\\
0,&r_{i}=0.\\
\end{cases}
\end{equation}

\begin{figure}[!h]
\centering
\includegraphics[width=3in]{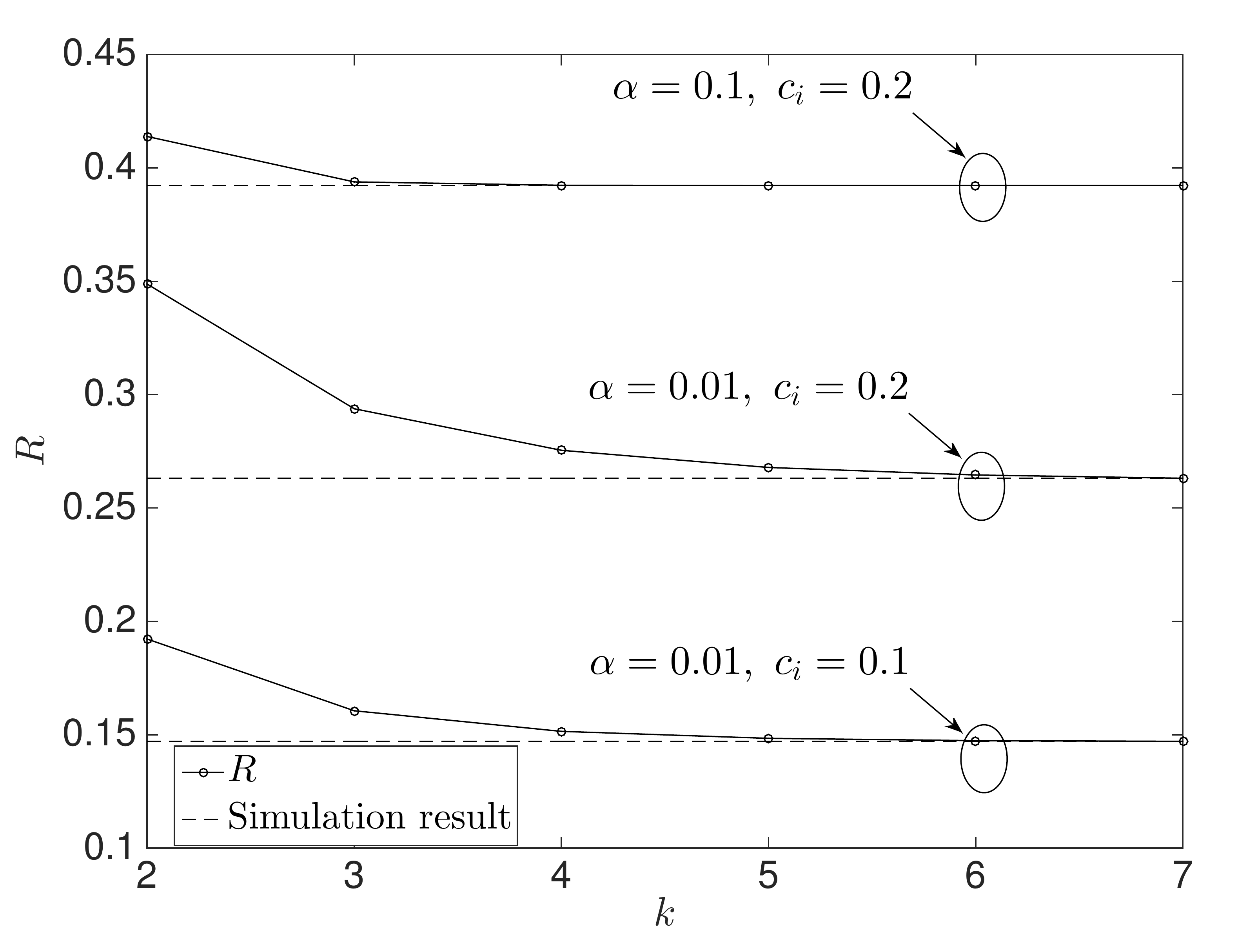}
\caption{The blockchain capacity of five miners that are connected with each other in a complete graph. The identical mining probabilities, $c_{i}$, and non-partition probability, i.e., $\alpha = a_{ij}$ are given in the figure. The dash lines are the simulation results running over $10^{6}$ timeslots.}
\label{fig_valInfiniteChain}
\end{figure}

Fig.~\ref{fig_valInfiniteChain} validates the accuracy of the approximation Markov model. There are five miners with the identical block mining rate $c_{i}$ and connected with each other in a complete graph. The synchronization success rates of all the non-ideal links are set to $\alpha$.
We can see that the capacity obtained from the approximated Markov model converges to the simulation results with the enlarging status space, i.e., increasing $k$. 
The convergence slows down with the decreasing transmission success rate and/or the increasing block mining rate.
This is because the decreasing transmission success rate and the increasing mining rate can increase the difference across the blockchain network, which needs large status spaces, i.e., large $k$, to capture the inconsistency.

\section{Capacity of two miners under eventual consistency}
\label{sec_captwominers}
\begin{figure}[!h]
\centering
\includegraphics[width=2.8in]{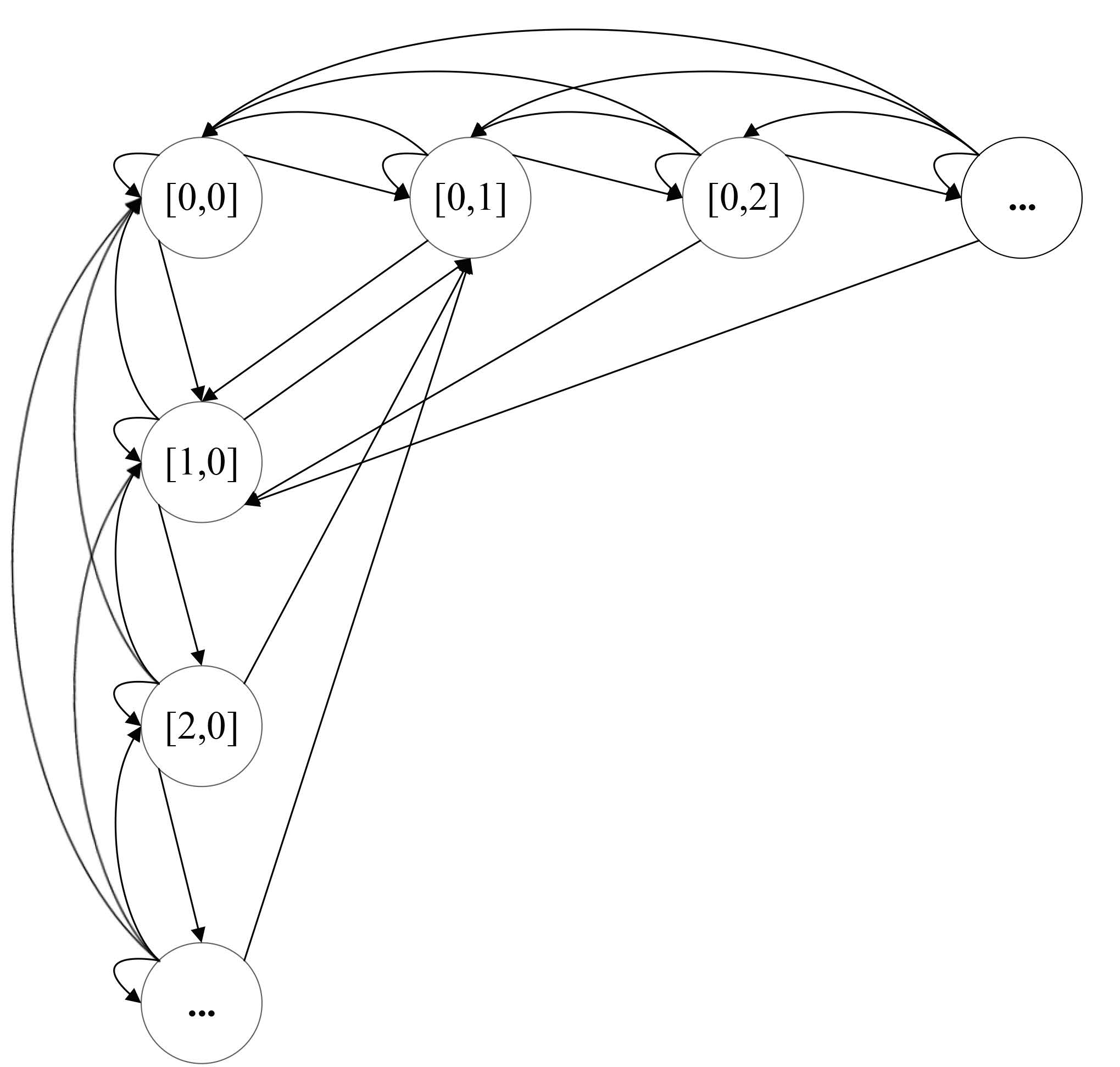}
\caption{State transition diagram of the two-dimensional EDTMC model of the blockchain growth between two miners connected by non-ideal links.}
\label{fig_twoClustersStatesDemoV2}
\end{figure}

In this section, we analyze a special case of the EDTMC model with two participating miners, where the closed-form expression for the eventual consistency capacity of a blockchain can be derived under unreliable connections between the miners, by employing queuing theory~\cite{kleinrock1976queueing}.
In this case, the EDTMC consists of two dimensions, as shown in Fig. \ref{fig_twoClustersStatesDemoV2}.

The state transitions among states $[i,0]$ (or $[0,j]$) can be interpreted as a queueing process with queue renovation or queue flushing.
From queueing theory, a queue renovation~\cite{kreinin1997queueing} or flushing~\cite{TOWSLEY1991353} process defines a process where a queue, e.g., M/M/1 queue~\cite{kleinrock1976queueing}, can be emptied at a predefined renovation probability per timeslot. States $[i,0]$ ($i\ge 0$) and $[0,j]$ ($j\ge 0$) can be visualized as two queues, of which the average arrival rates are $c_1 (1-c_2)$ and $(1-c_1) c_2$ and the average service rates are $(1-c_1) c_2$ and $c_1 (1-c_2)$, and the queue renovation probabilities~\cite{kreinin1997queueing} are $a_{12}$ and $a_{21}$, respectively.

The steady-state probability of state $[0,0]$, denoted by $\pi_{0,0}$, can be given in \eqref{equtspx001}, where the three terms account for three mutually exclusive scenarios.
(1) state $[0,0]$ can transit to itself if neither of the two miners mines a new block and the case that each miner generates a block independently.
(2) states $[i,0]$ and $[0,j]$ can transit to state $[0,0]$, if neither of the two miners mines a block or each miner produces a new block independently after successful synchronization.
(3) states $[1,0]$ and $[0,1]$ can transit to state $[0,0]$ if the miners fail to synchronize and the miner with the shorter chain mines a new block.
\begin{subequations}\label{equtspx00}\begin{align}
\pi_{0,0}=&(1-c_1-c_2+2 c_1 c_2)(\pi_{0,0}+a_{12}\sum_{i=1}^{{\infty}} \pi_{i,0}+ a_{21}\sum_{j=1}^{{\infty}} \pi_{0,j})\nonumber\\
&+(1-a_{12})(1-c_1) c_2 \pi_{1,0}+(1- a_{21})c_1 (1-c_2)\pi_{0,1};\label{equtspx001}\\
\pi_{1,0}=&c_1 (1-c_2) (\pi_{0,0}+a_{12}\sum_{i=1}^{{\infty}} \pi_{i,0}+ a_{21}\sum_{j=1}^{{\infty}} \pi_{0,j})\nonumber\\
&+(1-a_{12})(1-c_1-c_2+2 c_1 c_2) \pi_{1,0}+(1-a_{12})(1-c_1) c_2 \pi_{2,0};\label{equtspx002}\\
\pi_{i,0}=&(1-a_{12})c_1 (1-c_2) \pi_{i-1,0}+(1-a_{12})(1-c_1-c_2+2 c_1 c_2)\pi_{i,0}\nonumber\\
&+(1-a_{12})(1-c_1) c_2 \pi_{i+1,0},\text{ for}\ i\ge 2\label{equtspx003}.
\end{align}
\end{subequations}

Likewise, the steady-state probabilities of state $[i,0]$, i.e., $ \pi_{i,0}$, can be given in \eqref{equtspx002} and \eqref{equtspx003}.
Specifically, state $[0,0]$ can transit to state $[1,0]$ at the probability $c_1 (1-c_2)$ after the first miner produces a new block.
States $[i,0]$ and $[0,j]$ can transit to state $[1,0]$ with probabilities $a_{12} c_1 (1-c_2)$ and $ a_{21} c_1 (1-c_2)$, respectively, because only the first miner mines a new block after a successful synchronization.
States $[i,0]$ can also transit to itself in two cases: (1) neither of the two miners generates a block after a failed synchronization attempt, and the probability is $(1-a_{12})(1-c_1) (1-c_2)$; (2) each of the miners mines a block independently after a failed synchronization attempt, and the probability is $(1-a_{12}) c_1 c_2$.
State $[i-1,0]$ can transit to state $[i,0]$ with the probability $(1-a_{12})c_1 (1-c_2)$, if the first miner mines a new block after a failed synchronization attempt.
State $[i+1,0]$ can transfer to state $[i,0]$ with the probability $(1-a_{12})(1-c_1) c_2$, in the case that the second miner with the shorter chain mines a new block after a failed synchronization attempt.

The steady-state probabilities of state $[0,1]$ and state $[0,j]$, i.e., $\pi_{0,1}$ and $\pi_{0,j}$, can be obtained by switching the roles of the two miners in \eqref{equtspx002} and \eqref{equtspx003}. With the expressions of the steady-state probabilities, we can put forward the following theorem,

{
\theorem{The blockchain capacity of two miners, denoted by $R_2$, is given by
\label{the_1}
\begin{align*}
\begin{split}
R_{2} =&( c_1 + c_2 - c_1 c_2) \pi_{0,0}\\
&+( c_1 +a_{12} c_2 -a_{12}  c_1 c_2)\sum\nolimits_{i=1}^{\infty}\pi_{i,0}\\
&+( c_2 + a_{21} c_1 - a_{21}  c_1 c_2)\sum\nolimits_{i=1}^{\infty}\pi_{0,i}.
\end{split}
\end{align*}
}}
\begin{proof}
The proof of the theorem, and the numerical values of $\pi_{0,0}$, $\sum\nolimits_{i=1}^{\infty}\pi_{i,0}$ and $\sum\nolimits_{i=1}^{\infty}\pi_{0,i}$, are given in~\ref{app_1}.
\end{proof}

This theorem presents an accurate, closed-form expression for the blockchain capacity, which depends on the distribution of mining rates and non-ideal connections. This can avoid time-consuming simulations, e.g., in~\cite{Gervais:2016tu}.
The EDTMC in Section~\ref{sec_genmodel} can be hardly computed numerically due to its infinite statuses and could only be approximated by a Markov model with finite statuses, as done in Section~\ref{sec_genmodel}.
In contrast, the closed-form expression in this theorem is accurate and does not involve any approximation. This result does not rely on particular assumptions, such as unequal mining rates and bidirectional constant delays in~\cite{sompolinsky2015secure}, and therefore preserves adaptability.

\section{Capacity analysis under strong consistency}
\label{sec_stongcon}
Sections \ref{sec_genmodel} and \ref{sec_captwominers} can quantify the blockchain capacity, but cannot differentiate blocks mined by different miners in the globally endorsed blockchain. They cannot provide the probability at which a block has been admitted by all the miners. In this section, we develop a Fine-resolution DTMC (FDTMC) model from the perspective of \emph{strong consistency}, where both the number and the miners of blocks are traced.
The FDTMC model is extended from the proposed EDTMC model with increased dimensions recording the miners of individual blocks in the local chain of each miner.

\begin{figure}[!h]
\centering
\includegraphics[width=2.7in]{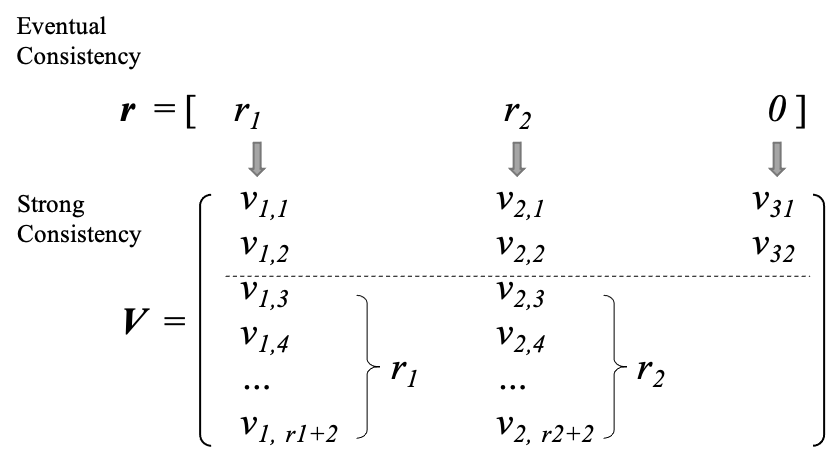}
\caption{A diagram of the connections between the proposed DTMC models under eventual consistency and strong consistency, where different local chains of three miners are illustrated with $r_3=0$.}
\label{fig_threeMinerscon}
\end{figure}

A state of $n$-dimensional FDTMC is denoted by $\mathbf V$, which is a set of $n$ vectors.
The $i$-th vector in $\mathbf V$, denoted by $\mathbf v_{i}$, records the miners of the blocks in the local blockchain of the $i$-th miner.
The $j$-th entry of $\mathbf v_{i}$, denoted by $v_{i,j}$, gives the index to the generating miner of the $j$-th block since the last state of strong consistency in the local chain of the $i$-th miner.
Fig. \ref{fig_threeMinerscon} provides an example of $\mathbf V$ with reference to $\bm r$ (a state of the EDTMC).
$\mathbf V=[\mathbf v_{1},\mathbf v_{2},\mathbf v_{3}]$, where the lengths of $\mathbf v_{1}$, $\mathbf v_{2}$, and $\mathbf v_{3}$ are $r_{1}+2$, $r_{2}+2$ and $2$, respectively; c.f., $\bm r=[r_{1},r_{2},0]$ in Section \ref{sec_genmodel}.
Nevertheless, the transition probability from $\mathbf V$ to $\mathbf V'$ can be evaluated in the same way as in Section \ref{sec_genmodel}, and therefore suppressed for brevity.

\begin{figure}[!h]
\centering
\includegraphics[width=2.5in]{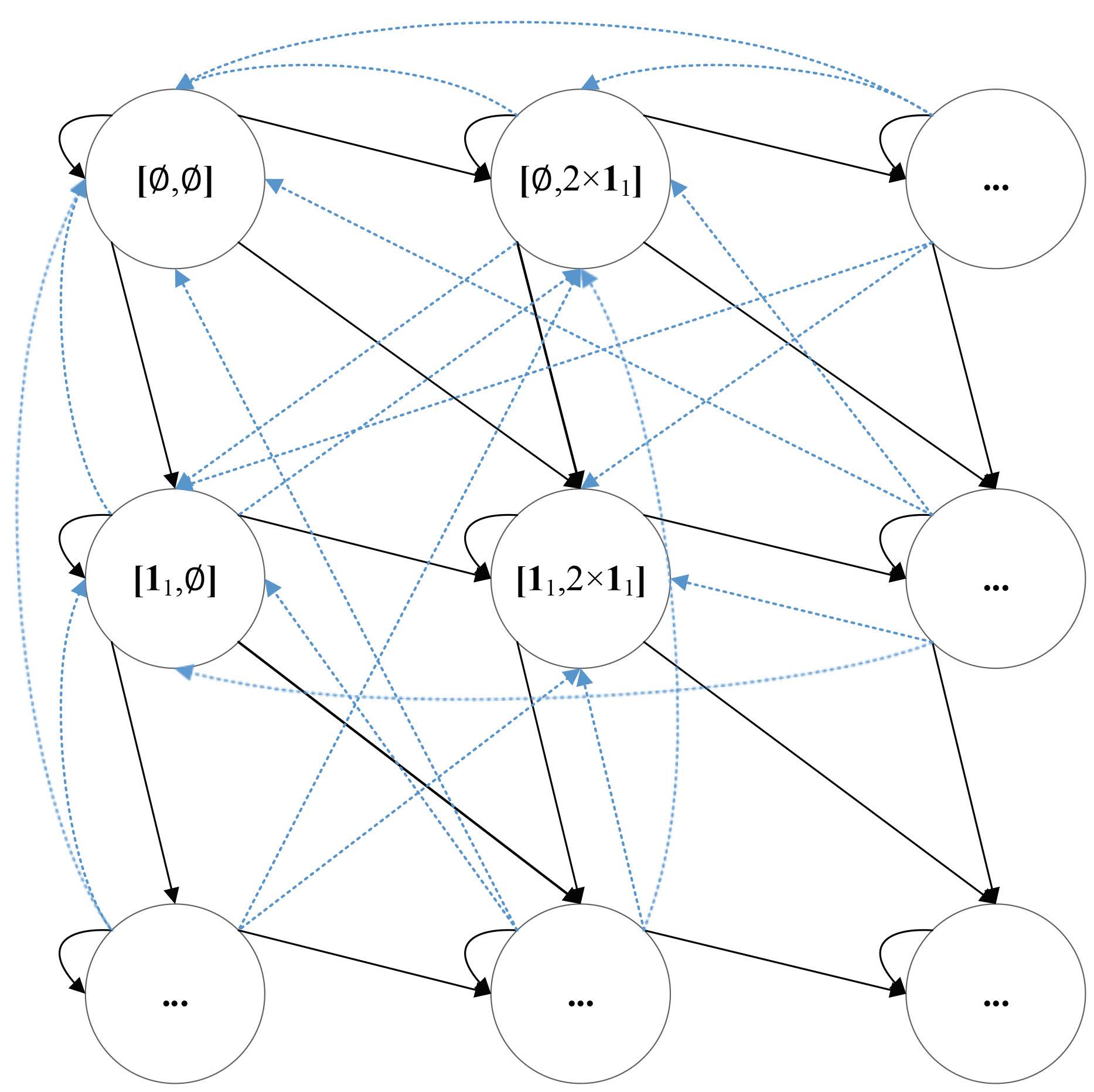}
\caption{States transition diagram of the FDTMC model of two miners connected by non-ideal links. The blue arrows are the state transitions with successful $\cal S$ phases, while the black arrows illustrate the state transitions with failed $\cal S$ phases.}
\label{fig_twoClustersStatesDemo}
\end{figure}

Fig.~\ref{fig_twoClustersStatesDemo} shows an example of the state transitions in a two-miner blockchain, where state $[\emptyset,\emptyset]$ is the state of strong consistency, i.e., the local chains of the two miners are exactly the same. State $[\mathbf 1_{i},2\times \mathbf 1_{j}]$ indicates that the first and the second miners have mined $i$ and $j$ different blocks in their local chains, respectively.
``$2$'' labels the $j$ blocks are mined at the second miner. 
This is because all states transit to $[\emptyset,\emptyset]$ once the two miners successfully exchange their knowledge on the longest chains.
The two inconsistent blocks can be mined by the two miners in the same timeslot, or in different timeslots without being properly synchronized between the miners.
With the transition probabilities, the steady-state probabilities of the FDTMC model with increased dimensions can be numerically and asymptotically evaluated to assess the strong consistency of the blockchain; see Section \ref{sec_genmodel}.

\label{equ_ETofTwoNodes}
In general, the analysis complexity of strong consistency can be exponentially higher than that of eventual consistency as studied in Section \ref{sec_genmodel}. This is due to the significantly increased dimensions of the states in the FDTMC model retaining fine details of the miners of individual blocks across the system, and a curse-of-dimensionality can occur~\cite{Indyk:1998:ANN:276698.276876}. Even in the case where two miners are connected by non-ideal links, the closed-form expression for the steady-state probabilities of the FDTMC with increased dimensions can be computationally prohibitive to deduce. 

\begin{figure}[!h]
\centering
\includegraphics[width=2.8in]{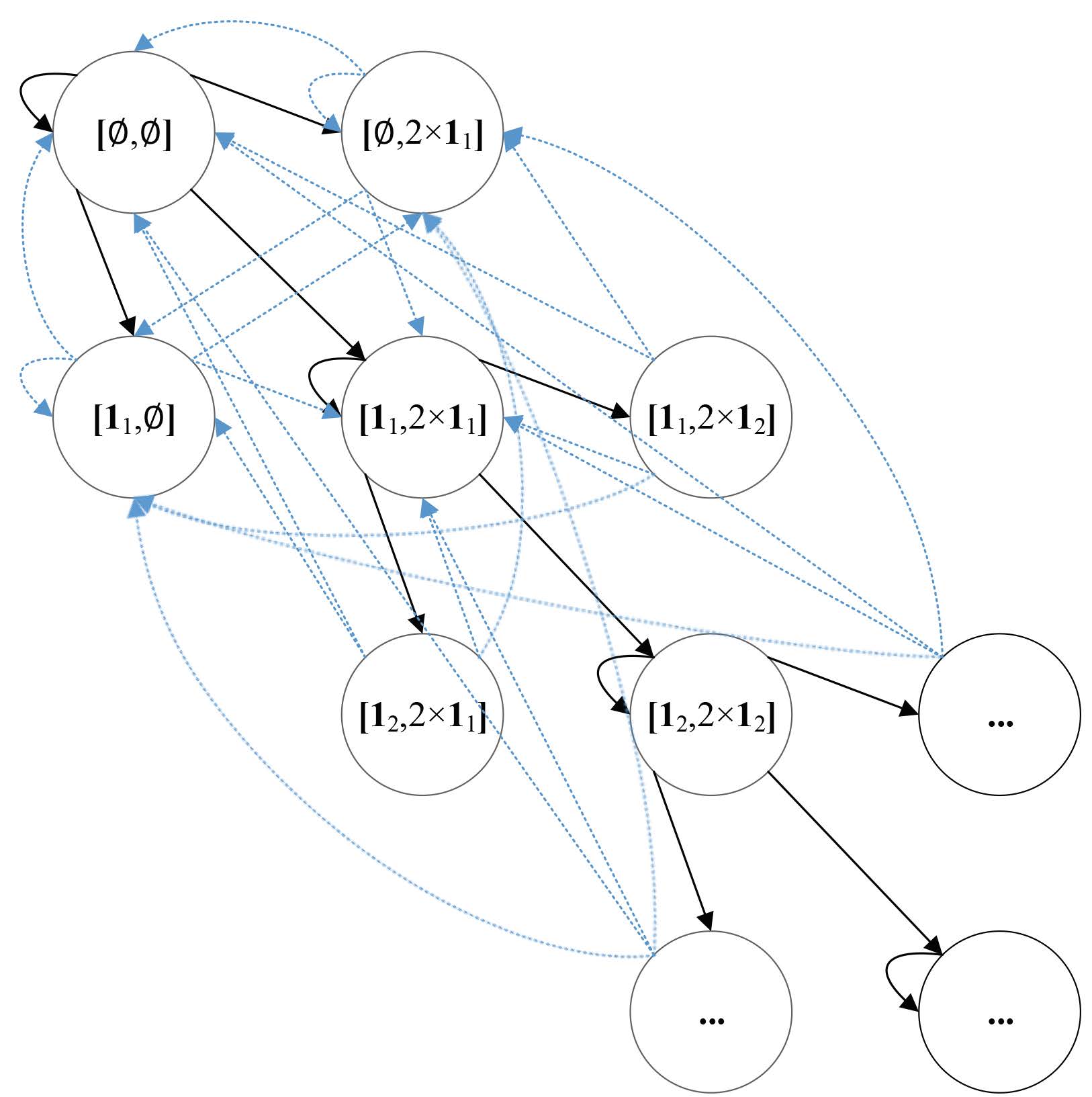}
\caption{An illustration of the state transition of the FDTMC model of two miners connected by ideal links. The blue arrows are the state transitions with successful $\cal S$ phases, while the black arrows illustrate the state transitions with failed $\cal S$ phases. }
\label{fig_twoNodesStatesDemo1beta1}
\end{figure}

In the ideal case where two miners are connected by error-free links, the closed-form expression for the capacity of blockchain with strong consistency can be derived. This is because the states can be substantially collapsed to reduce the dimensions of the FDTMC model, since valid transitions can only take place between states $[\mathbf 1_{i},2\times \mathbf 1_{j}]$ with $|i-j|\le1$, as illustrated in Fig.~\ref{fig_twoNodesStatesDemo1beta1}. With the error-free links, the two miners can successfully synchronize their local chains every timeslot. The difference between the lengths of the two blockchains cannot grow beyond one block.

As also shown in Fig. \ref{fig_twoNodesStatesDemo1beta1}, the non-diagonal states can transit to $[\emptyset,\emptyset]$ after synchronization.
The diagonal states can transit to the non-diagonal states after a timeslot, in which only a single block is mined from the two miners.
Let $\tau_{i,j}$ denote the (unique) steady-state probability of state $[\mathbf 1_{i},2\times \mathbf 1_{j}]$, and $\sum_{i}\sum_{j}\tau_{i,j}=1$. Thus, we have the following theorem,

{\theorem{After synchronization, the strong consistency probability between two miners connected by ideal links, denoted by $\eta$, is given by  
\label{theorem_2}
\begin{align*}
\begin{split}
\eta&=\tau_{0,0}+\sum_{i=1}(\tau_{i,i-1}+\tau_{i-1,i})=\frac{c_{1}+c_{2}-2c_{1}c_{2}}{c_{1}+c_{2}-c_{1}c_{2}}.
\end{split}
\end{align*}
$\eta$ is also the probability that a mined transaction is admitted by the two miners when the two miners have the strong consistency view of the globally endorsed blockchain.
}}

\begin{proof}
The proof of this theorem is given in \ref{app_2}. 
\end{proof}

The FDTMC model uses a large state space to trace all the unsynchronized blocks and their miners, while the EDTMC model developed in Sections~4 and~5 only captures the relative lengths of unsynchronized local chains. Thus, an EDTMC state covers multiple states in the FDTMC model. For example, $[0,0]$ in the EDTMC model covers states $[\mathbf 1_{i},2\times \mathbf 1_{i}],\ i\ge 0$ in the FDTMC model. All the states indicate that both local chains at the two miners have the same number of blocks. As a result, we can prove that $\pi_{0,0}=\sum_{i=0}^{\infty}, \tau_{i,i}$ in the case of $a_{12}=a_{21}=1$.
In addition to the blockchain capacity derived from the EDTMC model, the FDTMC model can provide the number of blocks mined by different miners in the globally endorsed blockchain, which can calculate the rewards for the miners, as stated in the following corollary

{
\corollary{
For two miners connected by ideal links, the ratios of the blocks mined by the first and second miner in the globally endorsed blockchain, denoted by $\gamma_{1}$ and $\gamma_{2}$, respectively, are given by
\label{corollary_1}
\begin{align*}
\begin{split}
\gamma_{1}=\frac{\sum_{i=1}^{\infty}(i\tau_{i,i-1})}{\bar R_{2}}=\frac{c_{1}(1-c_{2})}{c_{1}+c_{2}-2c_{1}c_{2}};\\
\gamma_{2}=\frac{\sum_{i=1}^{\infty}(i\tau_{i-1,i})}{\bar R_{2}}=\frac{c_{2}(1-c_{1})}{c_{1}+c_{2}-2c_{1}c_{2}}.
\end{split}
\end{align*}
}}

This is because the globally endorsed blockchain increases $i$ blocks from the first miner when state $[\emptyset,\emptyset]$ transits from states $[\mathbf 1_{i},2\times \mathbf 1_{i-1}]$; see $\sum_{i=1}^{\infty}(i\tau_{i,i-1})$ in the corollary.
Likewise, $\gamma_2$ can be obtained.

\section{Numerical Validation and Discussion}
\label{sec_simwithinlan}
We first evaluate the blockchain capacity of different numbers of miners, where the block mining rates meet $c_i=c_1\times q^{i-1}$ and $q$ is a scale factor. $c_{1}=\frac{1-q}{2(1-q^{n})}$, if $q\neq 1$; and $c_{1}=\frac{1}{2n}$, if $q=1$. We can tune the scale factor $q$ to simulate different mining rate distributions, e.g., identical mining rates across miners when $q=1$ and approximately centralized mining when $q=0.1$.

\begin{figure}[!h]
\centering
\includegraphics[width=3in]{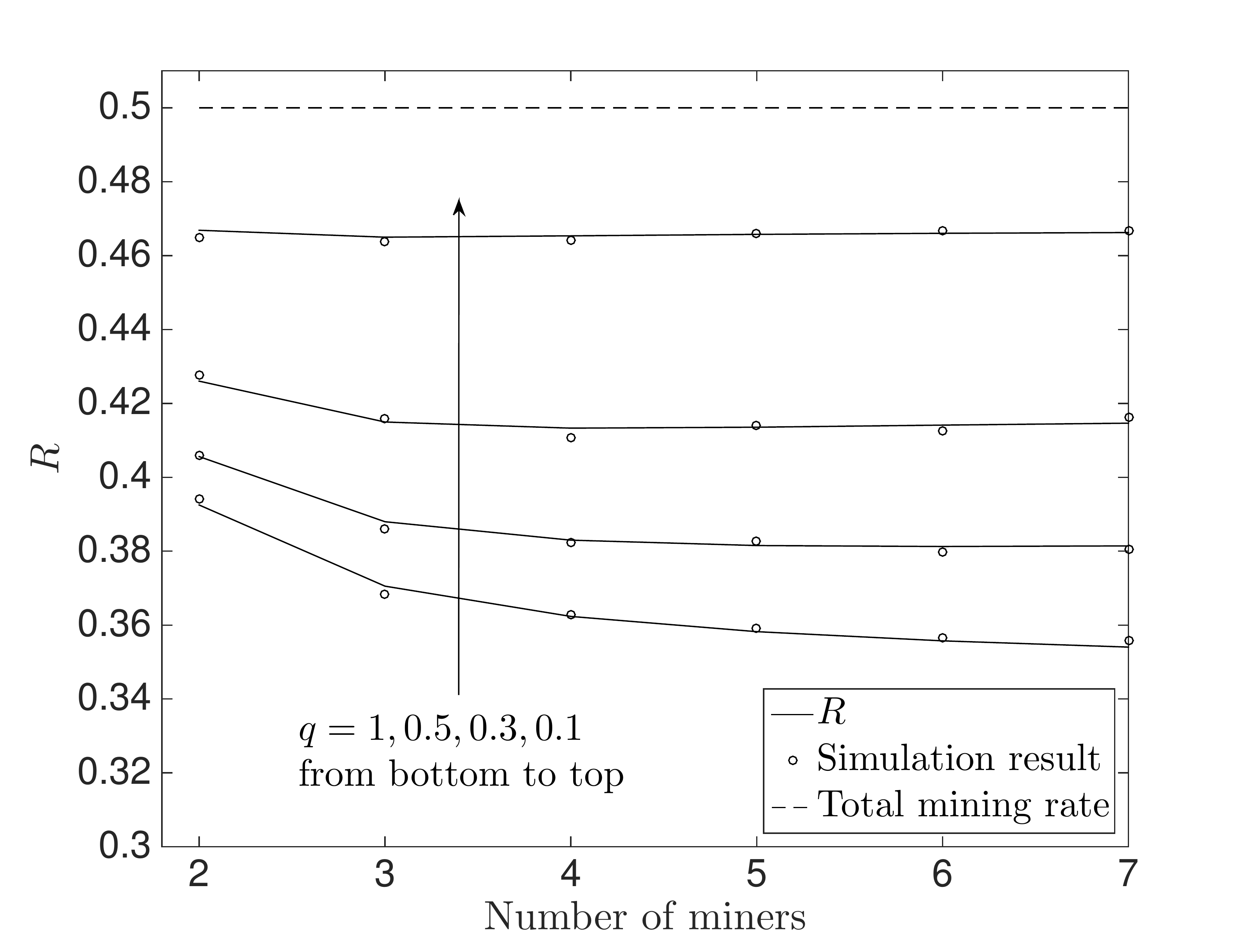}
\caption{Capacity of the public blockchain against the number of miners in a complete graph. The total mining rate is set to be 0.5, i.e., $\sum c_{i}=0.5$. $c_i=c_1\times q^{i-1}$, and $q$ is a scale factor. $c_{1}$ is set to $c_{1}=\frac{1-q}{2(1-q^{n})},\ q\neq 1$, and $c_{1}=\frac{1}{2n},\ q=1$.}
\label{fig_blockchainwithminers}
\end{figure}

Fig. \ref{fig_blockchainwithminers} shows $R$ against the number of miners under different scale factors based on the EDTMC in Section \ref{sec_genmodel}.
All the miners connect with each other in a complete graph.
The synchronization success rate is set to $\alpha=0.5$.
$a_{ij}=\alpha$ for $i\neq j$ and $a_{ii}=0$. The value of $q$ ranges from 0.1 to 1. The accumulated mining rate is $0.5$, i.e., $\sum_{i}c_{i}=0.5$, as shown by the reference line in the figure.
Each simulation result, indicated by a dot, takes an average of $2\times10^{6}$ timeslots.
As shown in the figure, blockchain capacity drops as the blockchain system is becoming more decentralized either with even mining rates or more miners.
The centralized mining, e.g., $q=0.1$, can achieve higher blockchain capacity than the decentralized mining, e.g., $q=0.3$.
The blockchain capacity is lower-bounded by that of homomorphic blockchain network where all the miners have identical mining rates.
We also notice that the blockchain capacity keeps stable for more than five miners.
This is because $c_{i}=\frac{(1-q)q^{{i-1}}}{2}$ is too small for $i>5$ and hardly contributes the block mining. 
As a result, the blockchain extension process can be dominated by the top-tier miners.

\begin{figure}[!h]
\centering
\includegraphics[width=3in]{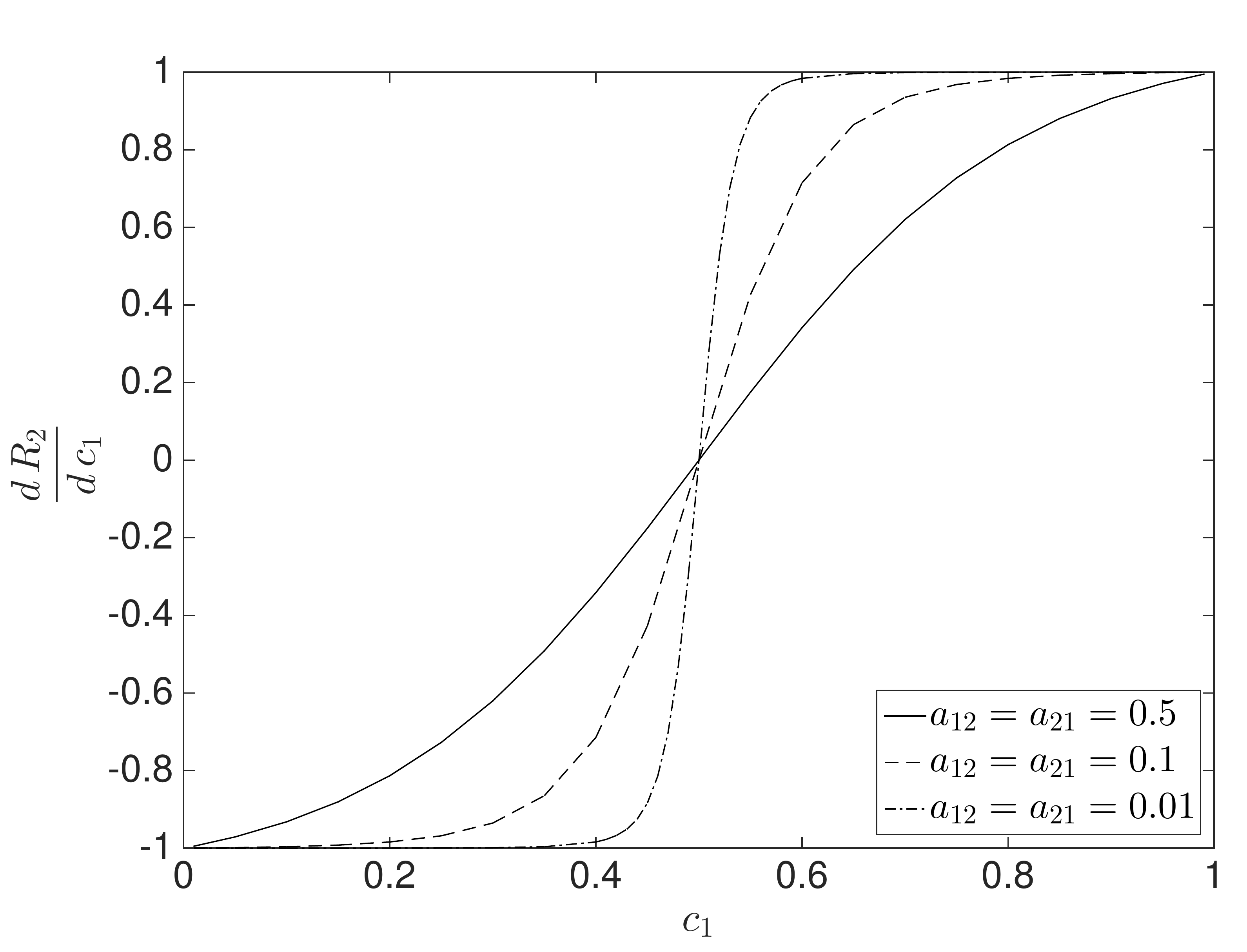}
\caption{$\frac{dR_{2}}{dc_{1}}$ with the growth of $c_{1}$, where $c_{2}=1-c_{1}$. Three transmission success rates, i.e., 0.5, 0.1 and 0.001, are considered.}
\label{fig_r2diff}
\end{figure}

Fig. \ref{fig_r2diff} plots $\frac{dR_{2}}{dc_{1}}$ with the growth of $c_{1}$, where we set $c_{2}=1-c_{1}$. The links are assumed to provide the same transmission success rates.
We can see that $R_{2}$ reaches the minimum value when $c_{1}=c_{2}$, because $\frac{dR_{2}}{dc_{1}}<0$ when $c_{1}<0.5<c_{2}$ and $\frac{dR_{2}}{dc_{1}}>0$ when $c_{1}>0.5>c_{2}$.
Particularly, the effect of the mining rate distribution can be analyzed with Theorem 1 and its derivative.
For example, in the case of $a_{12}=a_{21}=0.5$, $c_{1}+c_{2}=1$, we have $R_{2}=0.99$ when $c_{1}=0.01$.
The derivative of $R_2$ can be approximated as
\begin{equation}
\label{eq_der}
\begin{split}
\frac{d R_{2}}{d c_{1}}&\Bigg|_{c_{1} \to 0}\approx
\lim_{c_{1}\to 0}\Big(\frac{1.5 x^{1.5}+4 x^{0.5}+6}{x^{1.5} +x^{0.5} +  2}\\
&+ \frac{0.5}{0.25 x^{2.5} +  1.5 x^{1.5} +  0.25 x^{0.5}+2} \\
&- \frac{1}{ 0.5 x^{1.5} +0.5 x^{0.5} +1 } - \frac{3 x^{0.5}+4}{ x^{0.5} +  1}+O(c_{1})\Big)\\
&=-1,
\end{split}
\end{equation}
where $x=2 c_{1}^{2} - 2 c_{1} + 1$ and $\lim_{c_{1}\to 0 }x= 1$.

From the perspective of the blockchain growth, the two miners having the same block mining rates generate blocks at the same speed, which could result in conflicting blocks and slow down the growth of the main chain. This explains the observation in Fig. \ref{fig_blockchainwithminers} that the uniform block mining rates in a blockchain network would lead to the lowest blockchain capacity. We also can see that the blockchain capacity of the miners connected with poor links ($a_{12}=a_{21}=0.01$) decreases faster than the capacity of miners connected with good links ($a_{12}=a_{21}=0.5$), as $c_{1}$ grows from 0.05 to 0.5.

\begin{figure}[!h]
\centering
\includegraphics[width=3in]{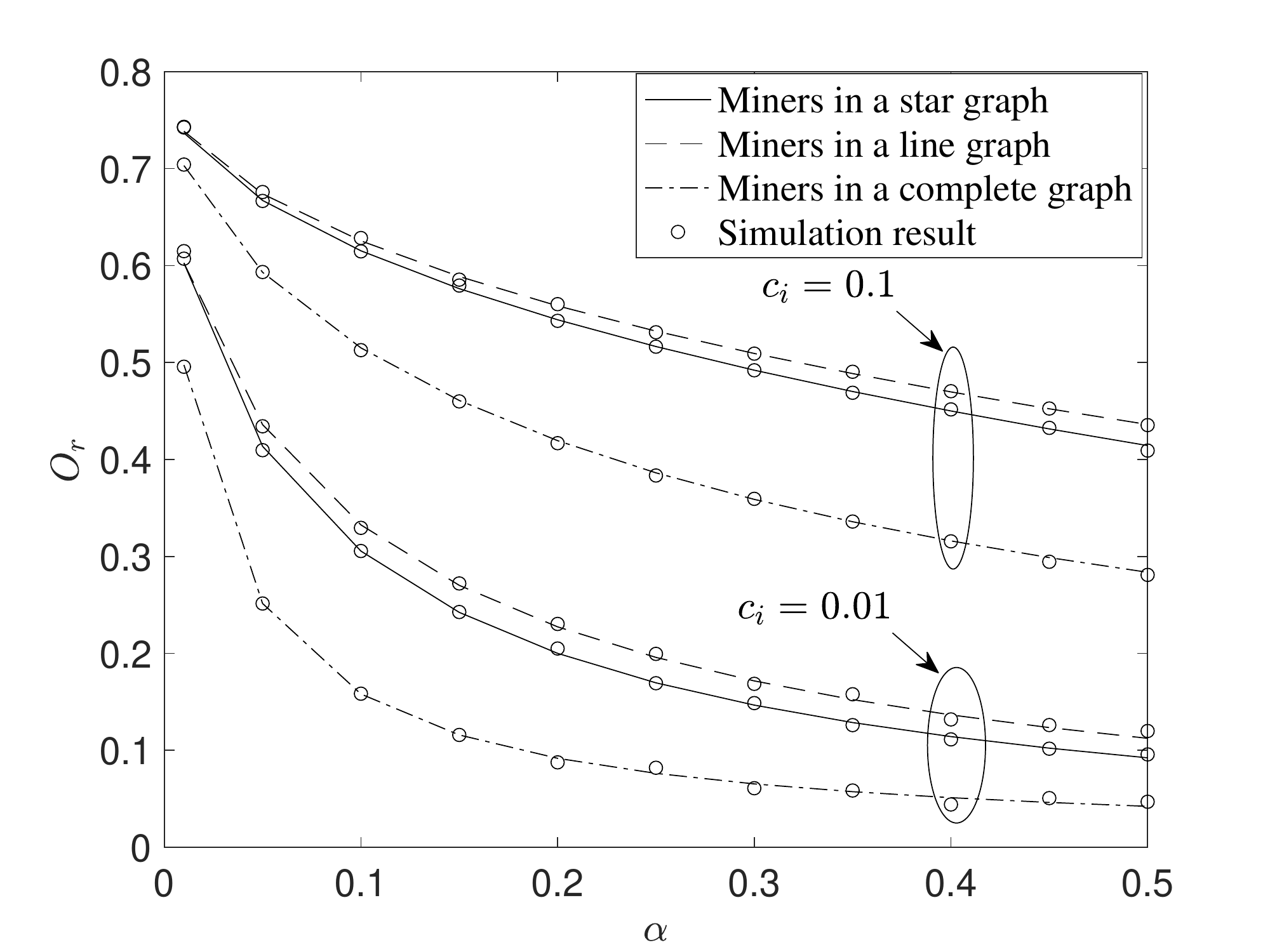}
\caption{The ratio of stale blocks, given by \eqref{equ_ratioStale}, with the growth of synchronization success rate $\alpha$, where 5-miners star, line and complete graphs are considered. The miners have identical mining rates, denoted by $c_{i}$.}
\label{fig_ratiostale}
\end{figure}

We evaluate the ratio of stale blocks using \eqref{equ_ratioStale} and \eqref{equ_rpiw}, against the growth of the synchronization success rate per link, denoted by $\alpha$, where the public blockchain system consists of five miners having the same block mining rate, i.e., $0.1$ and $0.01$, and the miners are connected in different graphs. Three types of graphs are considered, i.e., a line graph with diameter 4 and spectral radius 1.73, a star graph with diameter 2 and spectral radius 2, and a complete graph with diameter 1 and spectral radius 4.
We can see that $O_{r}$ declines with the growth of $\alpha$. Therefore, better link quality can help to reduce stale blocks.
Meanwhile, the graphs with small diameters or large spectral radiuses can achieve low $O_{r}$. 
Blockchains in the complete graph have lower $O_{r}$, especially when blocks are rapidly mined, i.e. $c_{i}=0.1$.
The star graph has the same number of edges as the line graph, but can achieve a lower stale block ratio.
We also see that blockchains can reduce the ratio of stale blocks to suppress attacks by slowing down the block mining, e.g., from $c_{i}=0.1$ to $c_{i}=0.01$, which scarifies blockchain capacity.

\begin{figure}[!h]
\centering
\includegraphics[width=3in]{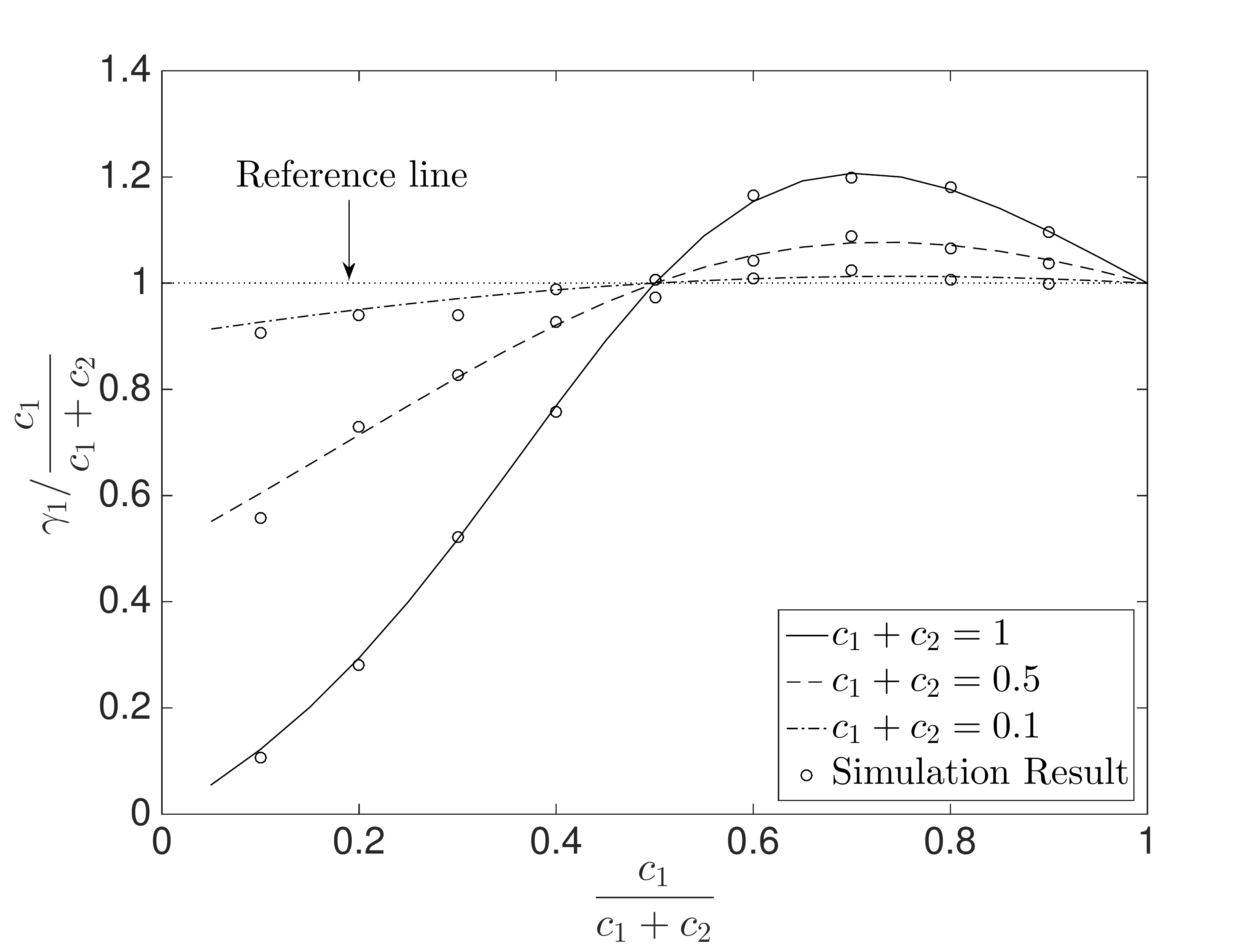}
\caption{The ratio of admitted blocks to computing power at the first miner, i.e., $\gamma_{1}/\frac{c_{1}}{c_{1}+c_{2}}$, with the growth of $\frac{c_{1}}{c_{1}+c_{2}}$}
\label{fig_gamma1}
\end{figure}

\begin{figure}[!ht]
\centering
\subfigure[The strong consistency probability after synchronization and the stale block ratio of the public blockchain.]{
\label{fig_twoMiners1}
\includegraphics[width=.45\columnwidth]{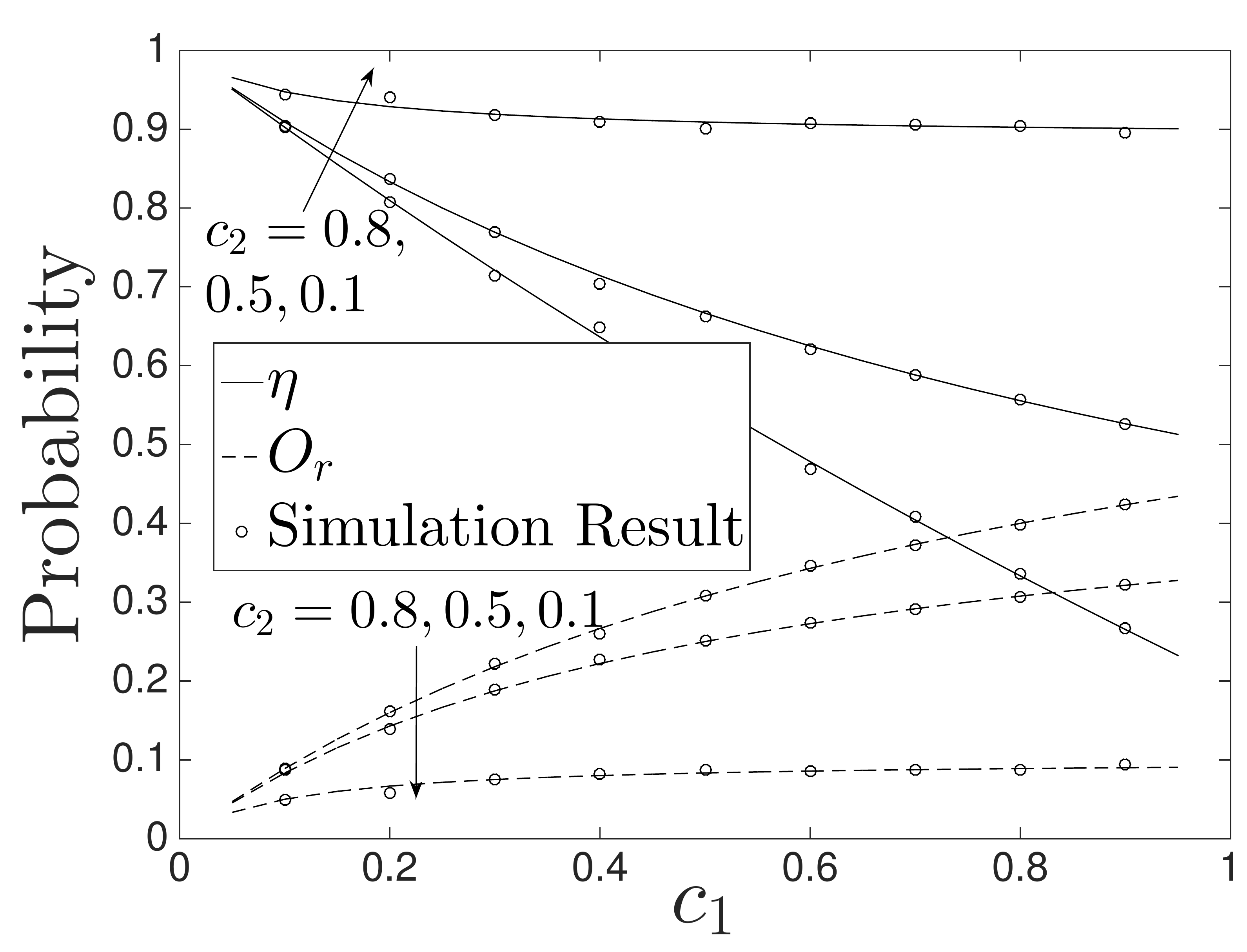}
}
\hspace{5mm}
\subfigure[Blockchain capacity contributed by two miners.]{
\label{fig_twoMiners2}
\includegraphics[width=.45\columnwidth]{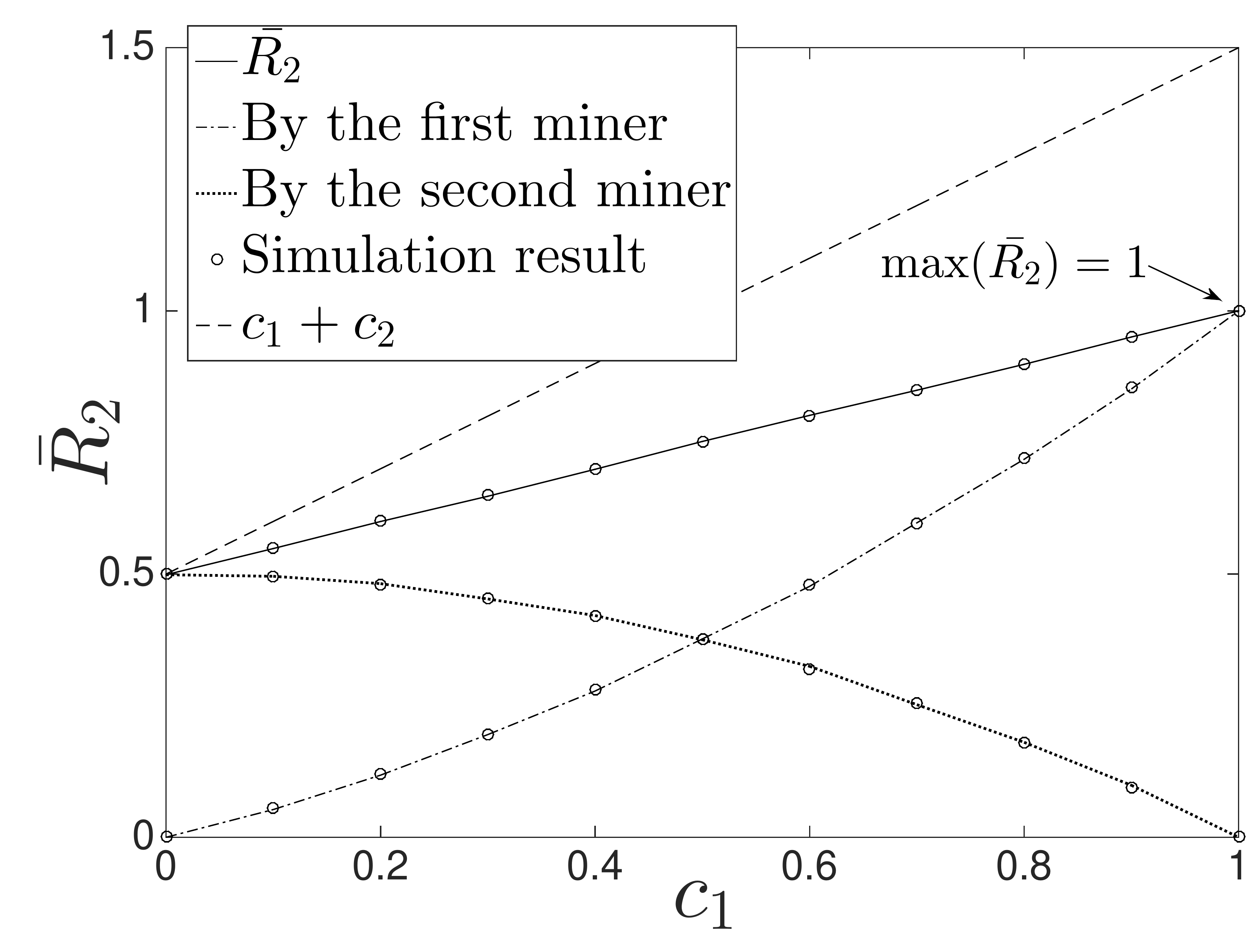}
}
\caption{The blockchain growth between two miners connected by ideal links.
}
\label{fig_twoMiners}
\end{figure}

We proceed to evaluate the strong consistency probability and the contributions of two miners to the blockchain growth based on Section \ref{sec_stongcon}.
Fig. \ref{fig_gamma1} plots the rate of successfully admitted blocks of the first miner to its computing power, i.e., $\gamma_{1}/\frac{c_{1}}{c_{1}+c_{2}}$, based on Corollary~\ref{corollary_1}, with the growth of $\frac{c_{1}}{c_{1}+c_{2}}$ in the ideal case where the link between the two miners is error-free. Different total block mining rates are considered.
We can see that the relative computing power (rather than the absolute computing power) of a miner has a strong impact on the number of successfully generated and admitted blocks in the final, eventually consistent blockchain.
In general, the miners can increase their computing power to improve their contributions to the final consistent blockchain.
However, the effect of increasing computing power is not linear.
As shown in the figure, the first miner achieves the maximum rate of admitted blocks to computing power when $\frac{c_{1}}{c_{1}+c_{2}}\approx 0.7$.
Fig. \ref{fig_twoMiners} plots the public blockchain growth between two miners connected by ideal error-free links.
To be specific, Fig. \ref{fig_twoMiners1} plots the strong consistency probability after synchronization, i.e., $\eta$ given in Theorem 2, and the stale block ratio, i.e., $O_{r}$ given by \eqref{equ_ratioStale}, where $c_{1}$ ranges with $[0,1]$ and $c_{2}$ is set to be 0.1, 0.5 and 0.8.
We can see that the strong consistency probability decreases with the increasing block mining rate, while the stale block ratio increases.
Fig. \ref{fig_twoMiners2} shows the contributions of each individual of the two miners to the consistent blockchain, where $c_{1}$ ranges with $[0,1]$ and $c_{2}$ is set to be 0.5.
We also see the blockchain capacity is upper bounded by 1, i.e., no more than one block per timeslot can be admitted by the main chain.

\begin{figure}[!ht]
\centering
\subfigure[$\bar{c}_{1}=0.0001$, $\bar{c}_{2}=0.0002$]{
\label{fig_comparison123v5_0}
\includegraphics[width=.46\columnwidth]{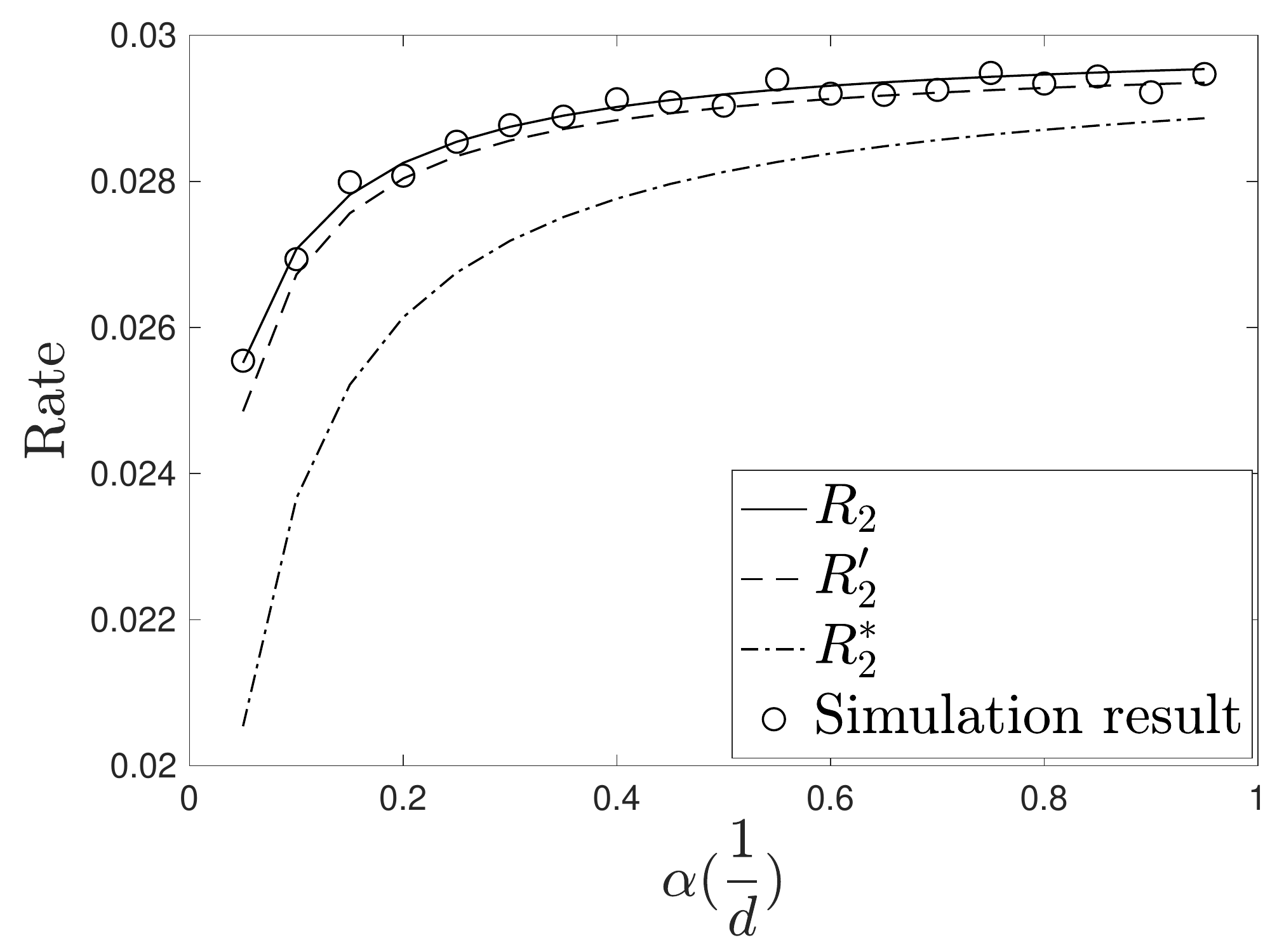}
}
\subfigure[$\bar{c}_{1}=0.001$, $\bar{c}_{2}=0.002$]{
\label{fig_comparison123v5_1}
\includegraphics[width=.46\columnwidth]{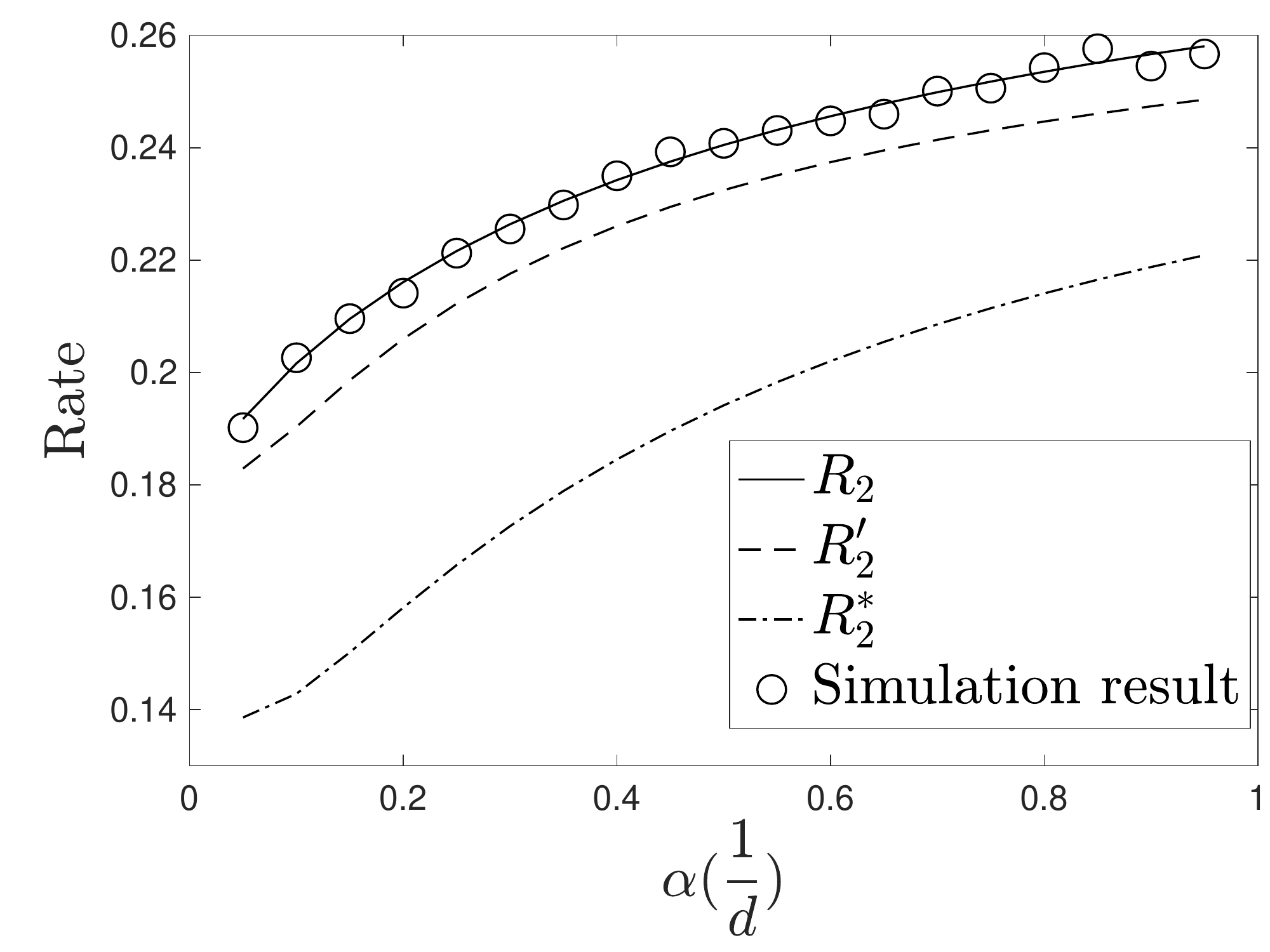}
}
\subfigure[$\bar{c}_{1}=0.025$, $\bar{c}_{2}=0.015$]{
\label{fig_comparison123v5_2}
\includegraphics[width=.46\columnwidth]{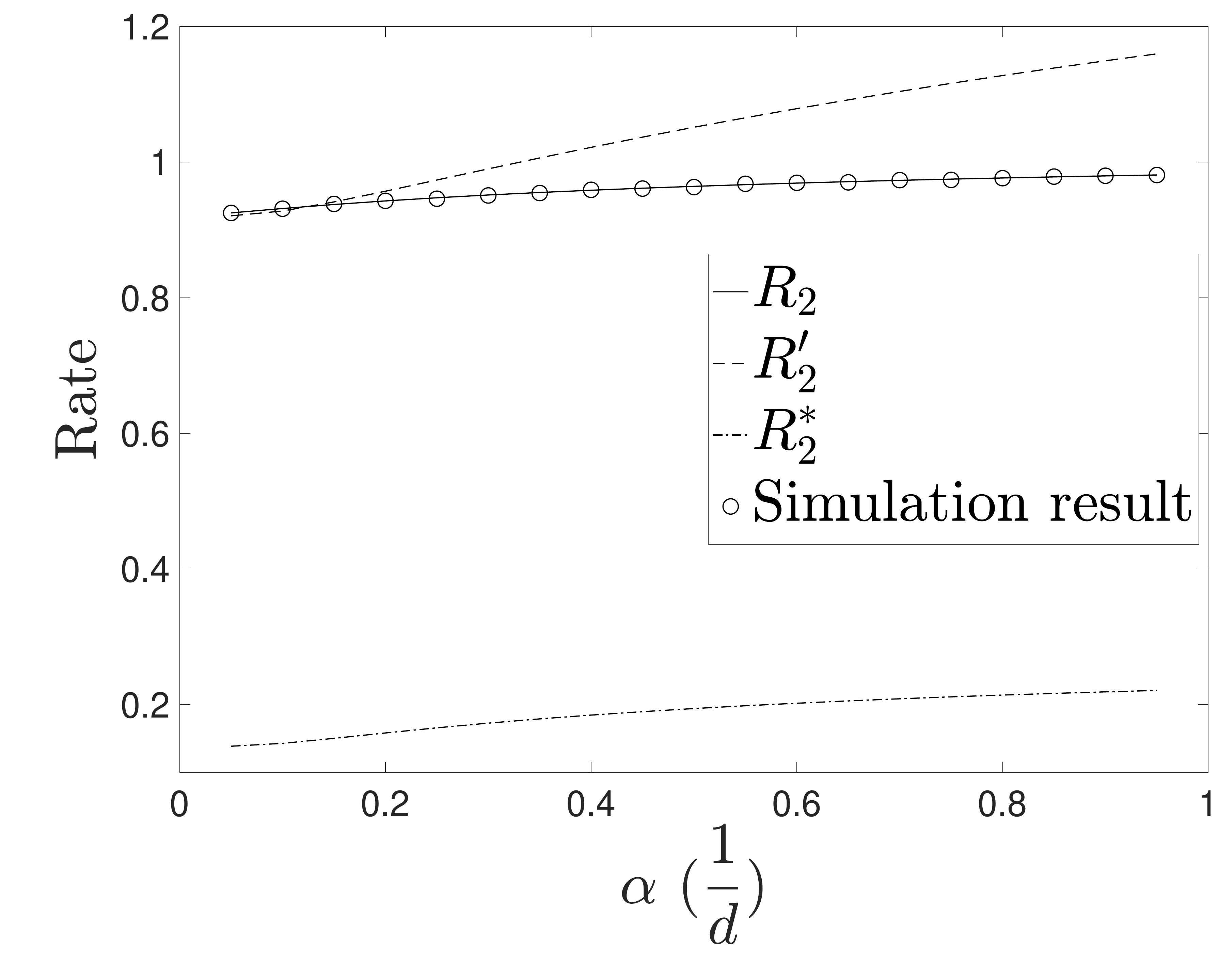}
}
\subfigure[$\bar{c}_{1}=0.02$, $\bar{c}_{2}=0.002$]{
\label{fig_comparison123v5_3}
\includegraphics[width=.46\columnwidth]{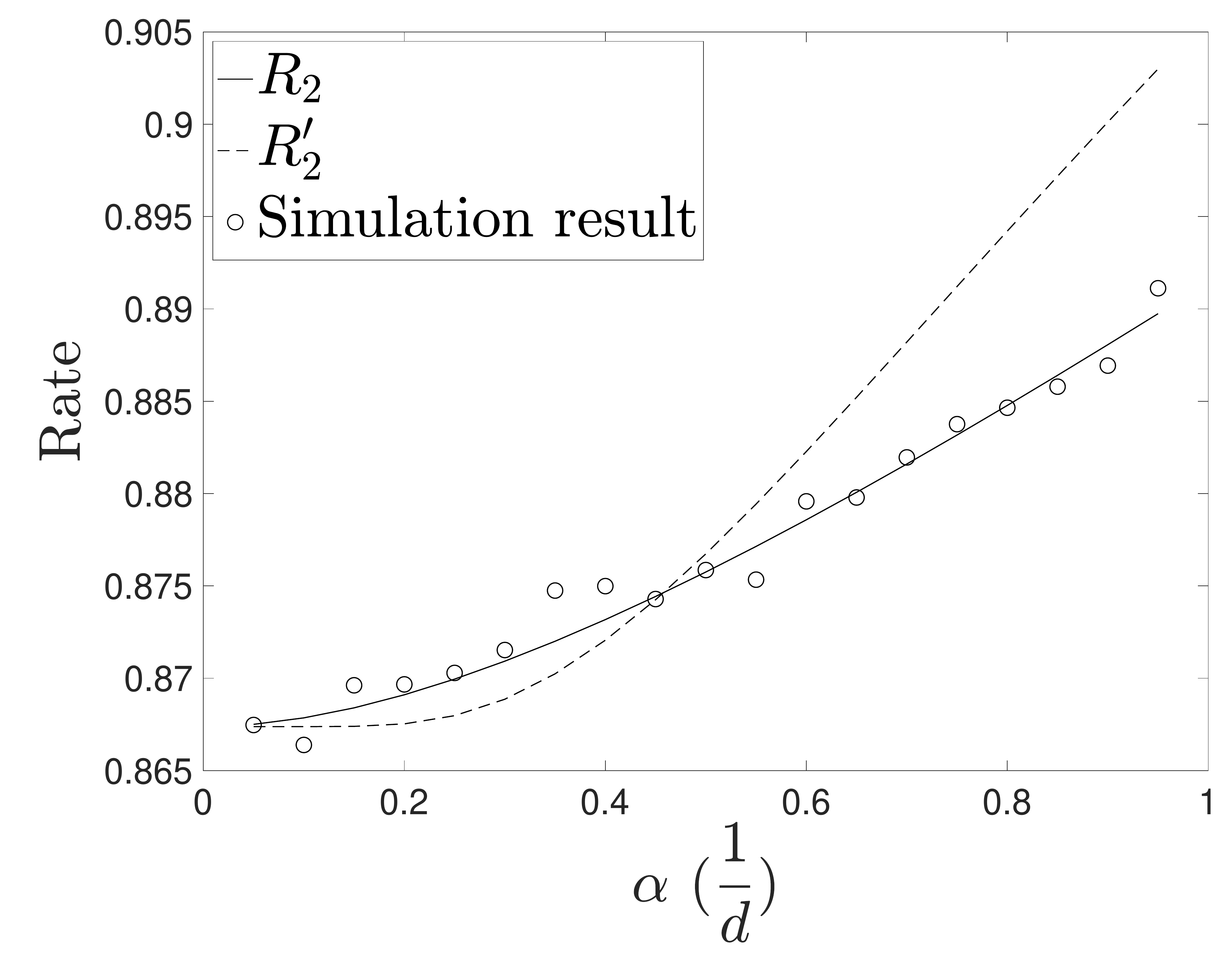}
}
\caption{The blockchain capacity between two 100-miner LANs, where $\bar{c}_{1}$ and $\bar{c}_{2}$ are the block mining rate per miner in the first LAN and the second LAN, respectively. 
The non-ideal links between the two LANs have the same transmission success probability $\alpha$.
$R_2$, $R_2'$ and $R_2^*$ are the theoretical blockchain growth rates from our work, \cite{sompolinsky2015secure} and \cite{8946275}, respectively. 
}
\label{fig_comparison123v5}
\end{figure}

We proceed to evaluate the capacity of a blockchain system across two 100-miner LANs based on the closed-form expression for the blockchain capacity in Fig.~\ref{fig_comparison123v5}.
Within each LAN, the miners are connected by error-free and delay-free links.
Between the LANs, the links are unreliable and can incur error, where we set $a_{12}=a_{21}=\alpha$.
Four configurations are considered. Configuration (a) corresponds to the case of low mining rates in the LANs.
Every miner in the first LAN has the block mining rate of $\bar{c}_{1}=0.0001$, while every miner in second LAN has the block mining rate of $\bar{c}_{2}=0.0002$.
Configuration (b) corresponds to the case of medium block mining rates, i.e., $\bar{c}_{1}=0.001$ and $\bar{c}_{2}=0.002$.
Configuration (c) corresponds to the case of a high block mining rate per miner in each of the LANs, i.e., $\bar{c}_{1}=0.025$ in the first LAN and $\bar{c}_{2}=0.015$ in the second LAN.
The last configuration (d) corresponds to the case where the two LANs have substantially different block mining rates per miner, i.e., $\bar{c}_{1}=0.02$ in the first LAN and $\bar{c}_{2}=0.002$ in the second LAN.
Each LAN can be interpreted as a miner with its capacity given by \eqref{equ_etg}, i.e., $c_1=1-(1-\bar c_1)^{100}$ and $c_2=1-(1-\bar c_2)^{100}$.

We also compare our work with blockchain models developed in~\cite{sompolinsky2015secure,8946275}. 
Our result from Theorem 1 is denoted by $R_2$. 
According to \cite[Theorem 9]{sompolinsky2015secure}, the blockchain growth rate of the two LANs, denoted by $R_2'$, can be given by 
\begin{equation}
\label{equ_tau}
R_2'=\frac{c_1^2e^{2c_1d}-c_2^2e^{2c_2d}}{c_1e^{2c_1d}-c_2e^{2c_2d}},
\end{equation}
where $e\approx  2.71828$ is Euler's number. $d$ is the block propagation delay between the LANs and can be converted from $\alpha$ by using $d=\frac{1}{\alpha}$ for comparison purpose.
According to~\cite[eq.~5]{8946275}, the blockchain growth rate, denoted by $R_2^*$, can be given by 
\begin{equation}
    R_2^*=\frac{c_1+c_2}{1+\Pr\{\text{fork}\}}=\frac{c_1+c_2}{1+1-e^{-d(c_1+c_2)}},
\end{equation}
where Pr\{fork\} is the blockchain fork probability and is given by with $1-e^{-d(c_1+c_2)}$.
The fork probability is under the assumption that a fork occurs if other blocks are mined before the currently proposed block is fully disseminated over the network.

Fig.~\ref{fig_comparison123v5} confirms that under all the four configurations, our analytical results, i.e., Theorem 1, provide good accuracy, evident from the simulation results. It is also shown in the figure that the blockchain capacity primarily depends on one of the LANs with higher capacity, especially when the LANs are connected with poor link qualities. An example is that the blockchain capacity is $0.868$, in the case where $\alpha=0.05$, $c_{1}=1-(1-0.02)^{100}=0.867$, and $c_{2}=1-(1-0.002)^{100}=0.181$, as shown in Fig.~\ref{fig_comparison123v5_3}.
The blockchain capacity can achieve its upper bound, i.e., $[1-(1-c_{1})(1-c_{2})]$, when the links between the LANs are error-free, i.e., $\alpha=1$.

The numerical values of $R_2'$ and $R_2^*$ mismatch the simulation results at different levels. It can be seen from Fig.~\ref{fig_comparison123v5_0} and Fig.~\ref{fig_comparison123v5_1} that $R_2'$ is slightly below the simulation results. This is because our model allows a chain of blocks to be synchronized with a successful transmission, while $R_2'$ assumes that all blocks have the same constant delay. The difference between our results and $R_2'$ is negligible for low mining rates, e.g., in Fig.~\ref{fig_comparison123v5_0}. 
Rapid block mining, e.g., under the setting of Fig.~\ref{fig_comparison123v5_2}, can beyond the assumption of $R_2'$~\cite[eq.~2]{sompolinsky2015secure} and result in non-negligible gaps. It is worth noting that $R_2'$ requires that two miners have different mining rates and only considers a bidirectional delay $d$ between the miners. 
Our model is general and able to capture the impact of uniform mining rates and asymmetric connections.    
It can be seen in Figs.~\ref{fig_comparison123v5_0} -- \ref{fig_comparison123v5_2} that $R_2^*$ underestimates the blockchain growth rates, and the gap between $R_2^*$ and the simulation results decreases with the increase of $\alpha$ or the decline of the mining rates (i.e., $c_1$ and $c_2$). The reason is that the model treats the case that the miners extend the longest chain before the end of block propagation delay~$d$ as a fork~\cite[eq.~5]{8946275} and therefore underestimates the blockchain growth rate. Nevertheless, this assumption simplifies the blockchain consensus analysis and is acceptable in the case of moderate fork rates, e.g., $\alpha=0.95$ in Fig.~\ref{fig_comparison123v5_0}.

\begin{figure}[!h]
\centering
\includegraphics[width=3in]{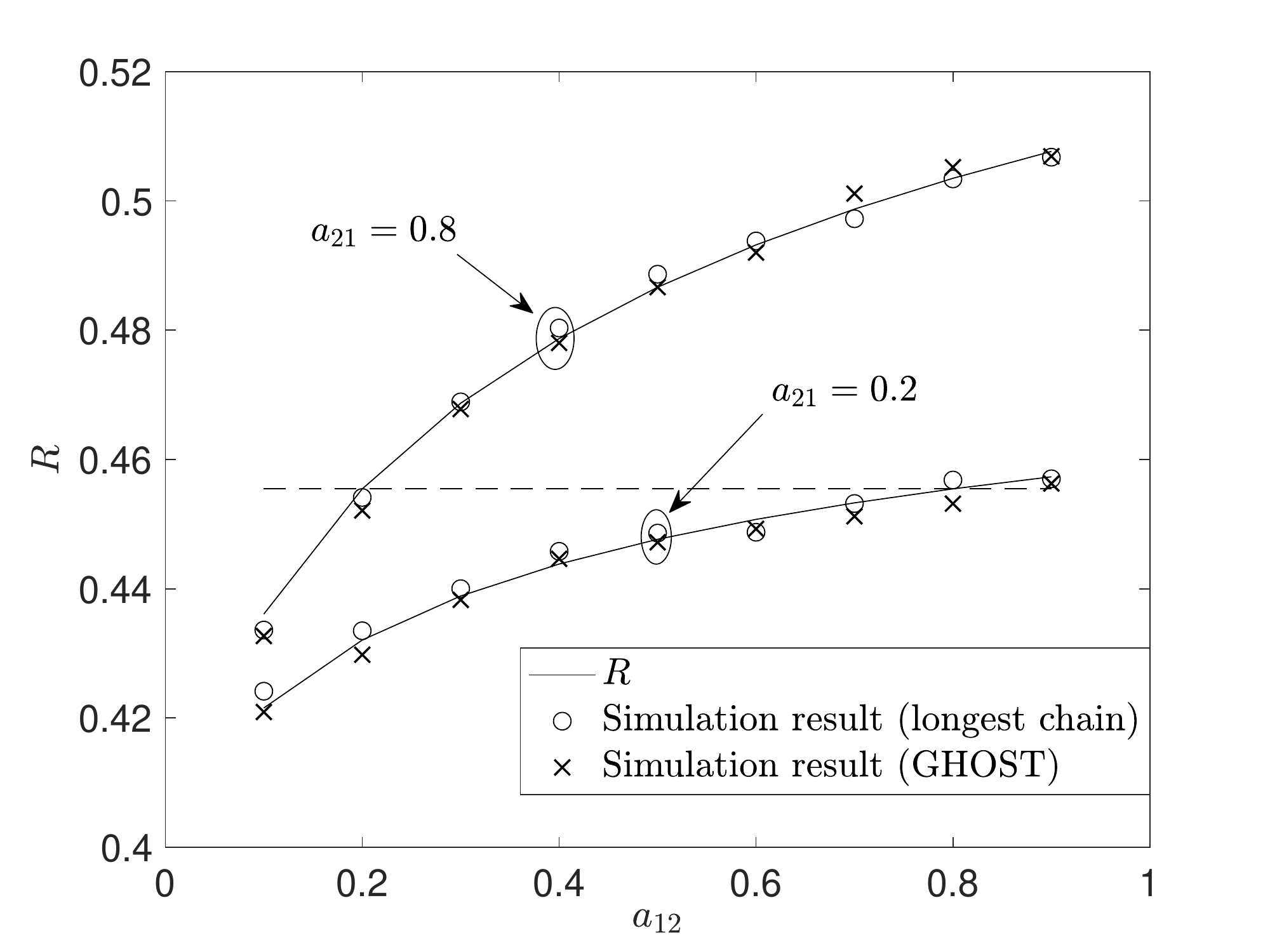}
\caption{The blockchain capacity with the growth of $a_{12}$, where two LANs are asymmetrically connected, i.e., $a_{12}\neq a_{21}$.}
\label{fig_catwolanchangingalpha}
\end{figure}

Fig. \ref{fig_catwolanchangingalpha} shows the blockchain capacity against the growth of $a_{12}$, where two LANs are connected by asymmetric links, i.e., $a_{12}\neq a_{21}$.
$a_{12}$ is set to range within [0.1,0.9], $ a_{21}$ is set to be 0.2 and 0.8 for the two solid curves.
$c_{1}=0.2$ and $c_{2}=0.4$. Every simulation dot in the figure is the average of $10^6$ timeslots.
We see that the blockchain capacity increases with the growth of $a_{12}$ and/or $ a_{21}$.
An interesting finding is that the switch of the roles between $a_{12}$ and $ a_{21}$ (or the first and the second LANs, equivalently) does not affect the blockchain capacity.
The blockchain capacity under the configuration of $a_{12}=0.2$ and $ a_{21}=0.8$ is equal to that under the configuration of $a_{12}=0.8$ and $ a_{21}=0.2$. Both are 0.455, as shown by the dash reference line.
We also see that the improvement of the lower of $a_{12}$ and $ a_{21}$ is more effective than the improvement of the higher. For example,
the configuration of $a_{12}=0.8$ and $ a_{21}=0.3$ can achieve higher blockchain capacity than the configuration of $a_{12}=0.9$ and $ a_{21}=0.2$ (0.4687 vs. 0.4573).
These findings indicate that it is practical to enhance the blockchain capacity by improving the lowest transmission success rate.

\begin{figure}[!h]
\centering
\includegraphics[width=3in]{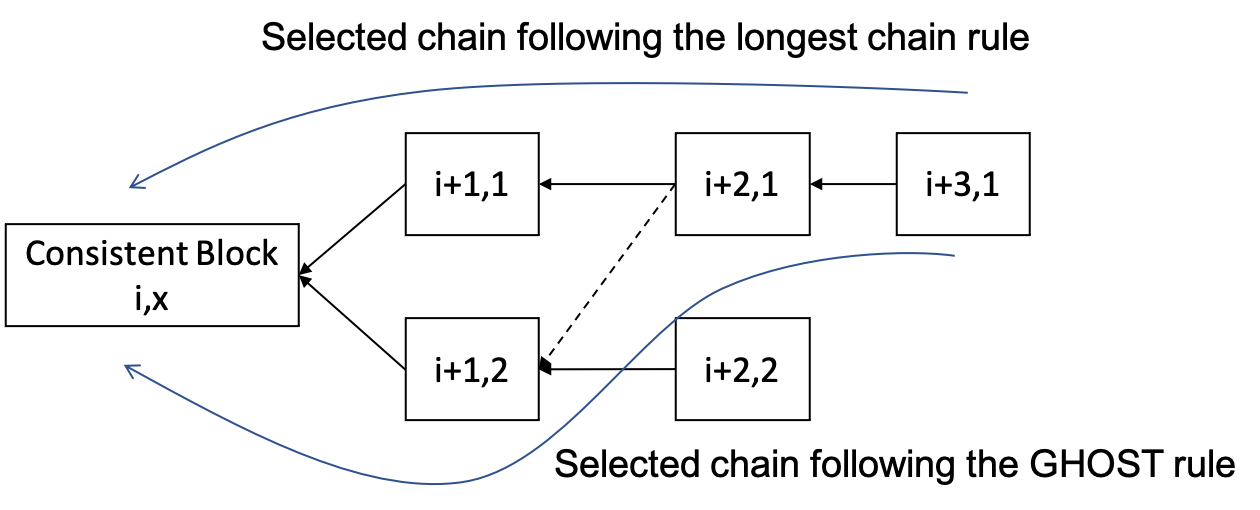}
\caption{Comparison between the chain selection rules of the longest chain and GHOST. Block $(i+k,l)$ indicates the block which is at the height of $(i+k)$ and mined by the $l$-th miner. The black arrows (both solid and dashed lines) indicate the hash links chaining the blocks. The solid arrows are valid under both rules, while the dashed arrow only exists under the GHOST protocol. The selected chains following the two rules are highlighted in blue. Block $(i+3,1)$ is the leaf block under both chain selection rules. Therefore, Theorem 1 applies to both rules. }
\label{fig_longestGhost}
\end{figure}

The proposed Markov models could be extended to capture other chain selection rules~\cite{buterin2017casper,sompolinsky2015secure,sompolinsky2016spectre} by defining Markov states to be the local blocks, which have not been globally endorsed, and accordingly updating the transition probabilities. For example, the states can be local direct acyclic graphs of blocks to analyze the chain extension in Ethereum that adopts the Greedy Heaviest Observed Subtree (GHOST) rule~\cite{buterin2017casper,sompolinsky2015secure} to select the globally endorsed chain, as opposed to the longest chain rule in Bitcoin. Theorem~\ref{the_1} can be directly applied to two Ethereum miners running the GHOST chain selection rule. This is because there are up to two blocks at the same height in the case of two miners, as discussed in~\cite{sompolinsky2015secure} and shown in Fig.~\ref{fig_longestGhost}. We can see that the leaf block selected by GHOST is also the latest block in the longest chain rule, although the two chains from the longest chain and GHOST rules can be different. Fig.~\ref{fig_catwolanchangingalpha} validates that the GHOST simulation results\footnote{We developed the GHOST experiment based on the GHOST simulation code from Ethereum (https://github.com/ethereum/economic-modeling/blob/master/ghost.py)}, denoted by ``$\times$'', are consistent with the theoretical $R$ and the simulation results with the longest chain rule. 

\section{Conclusion}
\label{sec_con}
In this paper, infinite-state Markov models were developed to capture inconsistent block generations or forks for a PoW based public blockchain.
The steady-state probabilities of the models were shown to be convergent and derived in closed-form, and revealed that the blockchain capacity can be analyzed in a structured way by sequentially evaluating the capacity of different network partitions in a recursive manner. The impact of the mining rates of individual miners and the link conditions between miners on the capacity was quantified. Confirmed by simulations, our models showed that the blockchain capacity depends on the distribution of the mining rates, and is lower-bounded by the case with uniform mining rates.

\section*{Acknowledgement}
This work was supported by BT Group plc through the Blockchain Enabled Inventory Management - A Pilot Project.

\appendix
\section{Proof of Theorem 1}
\label{app_1}
To analytically evaluate the blockchain capacity, we define the probability generation functions (PGFs) of $ \pi_{i,0}\ (i\ge 2)$ and $ \pi_{0,j}\ (j\ge 2)$, as given by
\begin{equation}
\begin{split}
\phi(z)&=\sum_{i=2}^{\infty} \pi_{i,0}z^{i-1},\ 0<z\le 1;\\
\psi(z)&=\sum_{j=2}^{\infty} \pi_{0,j}z^{j-1},\ 0<z\le 1.
\end{split}
\end{equation}
We have
\[
\phi(1)=\sum_{i=2}^{\infty} \pi_{i,0}\text{ and }\psi(1)=\sum_{j=2}^{\infty} \pi_{0,j}.
\]

Multiplying \eqref{equtspx003} with $z^{i}$ and then adding up over $i$,  we have
\begin{equation}
\label{equx20x101}
\begin{split}
&\big[1-(1-a_{12})((1-c_1) (1-c_2)+ c_1 c_2)\big]z\phi(z)\\
&\!=\!(1\!-\!a_{12})c_1 (1\!-\!c_2)z^{2} \pi_{1,0}\!+\!(1\!-\!a_{12})c_1 (1\!-\!c_2)z^{2}\phi(z)\\
&\!+\!(1-a_{12})(1-c_1) c_2\phi(z)\!-\!(1\!-\!a_{12})(1-c_1) c_2z \pi_{2,0},
\end{split}
\end{equation}
which can be rearranged as
\begin{equation}
\label{equx20x102}
\begin{split}
&\Big[(1-a_{12})c_1 (1-c_2)z^2+(1-a_{12})(1-c_1) c_2\\
&-[1-(1-a_{12})((1-c_1) (1-c_2)+ c_1 c_2)]z\Big]\phi(z)\\
&=(1\!-\!a_{12})(1\!-\!c_1) c_2z \pi_{2,0}\!-\!(1\!-\!a_{12})c_1 (1\!-\!c_2)z^2 \pi_{1,0}.
\end{split}
\end{equation}
By plugging different values of $z$, both $\pi_{2,0}$ and $\phi(1)=\sum_{i=2}^{\infty} \pi_{i,0}$ can be written as functions of $ \pi_{1,0}$.
We first, set both sides of \eqref{equx20x102} to 0, $z\neq 0$ and $\phi(z)\neq 0$.
We have $f(z)=(1-a_{12})c_1 (1-c_2)z^2-\big[1-(1-a_{12})((1-c_1) (1-c_2)+ c_1 c_2)\big]z+(1-a_{12})(1-c_1) c_2=0$. Because $f(0)=(1-a_{12})(1-c_1) c_2>0$, $f(1)=-a_{12}<0$ and $f(+\infty)>0$, $f(z)=0$ has a single root within $(0,1)$, denoted by $z_1$. The small root,  $z_1$, is given in \eqref{equ_z1} by applying the quadratic formula.

Let $z=z_1$ in the right-hand side (RHS) of \eqref{equx20x102}. Both the RHS and the left-hand side (LHS) become $0$. We have
\begin{equation}
\label{equx20}
 \pi_{2,0}=\frac{c_1 (1-c_2)z_1}{(1-c_1) c_2} \pi_{1,0}.
\end{equation}

Next, $\phi(1)$ can be obtained by substituting \eqref{equx20} and $z=1$ in \eqref{equx20x102}. Both sides of \eqref{equx20x102} provide the same non-zero value. $\phi(1)$ can be given by
\begin{equation}
\label{equx10}
\begin{split}
\phi(1)=\frac{(1-a_{12})c_1 (1-c_2)(1-z_1)}{a_{12}} \pi_{1,0}.
\end{split}
\end{equation}

The equation about $\psi(z)$ can be obtained by switching the roles of the two miners and, subsequently, $\phi(z)$ and $\psi(z)$ in \eqref{equx20x102}, and can be given by
\begin{equation}
\label{equx02x01}
\begin{split}
&\Big[(1- a_{21})(1-c_1) c_2z^2+(1- a_{21})c_1 (1-c_2)\\
&-[1-(1- a_{21})((1-c_1) (1-c_2)+ c_1 c_2)]z\Big]\psi(z)\\
&=\!(1\!-\! a_{21})c_1 (1\!-\!c_2)z \pi_{0,2}\!-\!(1\!-\! a_{21})(1\!-\!c_1) c_2z^2 \pi_{0,1}.
\end{split}
\end{equation}
The quadratic equation $g(z)=(1- a_{21})(1-c_1) c_2z^2-[1-(1- a_{21})((1-c_1) (1-c_2)+ c_1 c_2)]z+(1- a_{21})c_1 (1-c_2)=0$ has only a single root within $(0,1)$, denoted by $z_3$. This is because $g(0)=(1- a_{21})c_1 (1-c_2)>0$, $g(1)=- a_{21}<0$ and $g(+\infty)>0$.
$z_3$ can be obtained by switching the roles of the two miners in \eqref{equ_z1}.

Let $z=z_3$ in the RHS of \eqref{equx02x01}. Both sides of \eqref{equx02x01} become $0$. We have
\begin{equation}
\label{equx02}
 \pi_{0,2}=\frac{(1-c_1) c_2z_3}{c_1 (1-c_2)} \pi_{0,1}.
\end{equation}

By plugging \eqref{equx02} and $z=1$ in \eqref{equx02x01}, $\psi(1)$ can be given by
\begin{equation}
\label{equx01}
\begin{split}
\psi(1)=\frac{(1- a_{21})(1-c_1) c_2(1-z_3)}{ a_{21}} \pi_{0,1}.
\end{split}
\end{equation}

\begin{equation}
\label{equ_z1}
\begin{split}
z_1=&\Big(2\left(1-a_{12}\right)c_1\left(1-c_2\right)\Big)^{-1}\\
&\times\bigg(1-\left(1-a_{12}\right)\left(1- c_1 - c_2 +2 c_1 c_2\right)\\
&\qquad-\Big(\big(1-\left(1-a_{12}\right)\left(1- c_1 - c_2 +2 c_1 c_2\right)\big)^2\\
&\qquad\qquad-4\left(1-a_{12}\right)^2 c_1 c_2\left(1-c_1\right) \left(1-c_2\right)\Big)^\frac{1}{2}\bigg).
\end{split}
\end{equation}
\begin{subequations}
\label{equ_2difirst}
\begin{align}
 \pi_{1,0}=&\Big(c_1 (1-c_2)\big(e a_{21}  c_2 (1- c_1 )+d\big)\Big)\nonumber\\
 &\times\Big(\big( f + d c_1 (1-c_2)\big)\big( g +e c_2 (1- c_1 )\big)\nonumber\\
 &\qquad- d e c_1 c_2(1-c_1) (1-c_2)(1-a_{12})(1- a_{21})\Big)^{-1};\label{equ_2difirst1}\\
 \pi_{0,0}&=1-d \pi_{1,0}-e \pi_{0,1}\label{equ_2difirst2}.
\end{align}
\end{subequations}

Substituting \eqref{equtspx00}, \eqref{equx20},  \eqref{equx10},  \eqref{equx02},  \eqref{equx01} and $ \pi_{0,1}$ into $ \pi_{0,0}+ \pi_{1,0}+\phi(1)+ \pi_{0,1}+\psi(1)=1$, we can obtain $ \pi_{1,0}$ and $ \pi_{0,0}$, as given in \eqref{equ_2difirst}, where $ d =1+\frac{(1-a_{12})c_1 (1-c_2)(1-z_1)}{a_{12}}$,
$ e =1+\frac{(1- a_{21})(1-c_1) c_2(1-z_3)}{ a_{21}}$,
$ f =1-(1-a_{12})[(1-c_1) (1-c_2)+ c_1 c_2+c_1 (1-c_2)z_1]-da_{12} c_1 (1-c_2)$,
and $ g =1-(1- a_{21})[(1-c_1) (1-c_2)+ c_1 c_2+ c_2 (1-c_{1})z_3]-e a_{21}  c_2 (1- c_1 )$.
$ \pi_{0,1}$ can be obtained by switching the roles of the first and the second miners, i.e., switching $d$ and $e$, and switching $f$ and $g$ in \eqref{equ_2difirst1}.
In turn, $\phi(1)$ and $\psi(1)$ can also be obtained.

According to \eqref{equ_rpiw} and \eqref{sec_ww}, the capacity of the two miners, denoted by $R_{2}$, can finally be given by
\begin{subequations}
\label{equ_throughputoftwoclusterswithcom}
\begin{align}
R_{2} =&( c_1 + c_2 - c_1 c_2) \pi_{0,0}\label{equ_mergeE1}\\
&+( c_1 +a_{12} c_2 -a_{12}  c_1 c_2)( \pi_{1,0}+\phi(1))\label{equ_mergeE2}\\
&+( c_2 + a_{21} c_1 - a_{21}  c_1 c_2)( \pi_{0,1}+\psi(1))\label{equ_mergeE3},
\end{align}
\end{subequations}
where \eqref{equ_mergeE1} is because the longer blockchain extends a single block if no less than one block is mined in the state $[0,0]$. \eqref{equ_mergeE2} is the probability that the first miner generates a new block per timeslot or only the second miner generates a new block after successful synchronization in state $[i,0],\ i>0$. \eqref{equ_mergeE3} is the probability that the second miner produces a new block per timeslot or only the first miner generates a new block after successful synchronization in state $[0,j],\ j>0$. It is clear that $R_2$ depends on the error probabilities of the link and the delay of the links.

In a special case with error-free links, $a_{12}= a_{21}=1$, \eqref{equ_throughputoftwoclusterswithcom} can be rewritten as
\begin{equation}
\label{equ_barr2}
\begin{split}
\bar R_{2}&=(c_{1}\!+\!c_{2}\!-\!c_{1}c_{2})( \pi_{0,0}\!+\! \pi_{1,0}\!+\!\phi(1)\!+\! \pi_{0,1}\!+\!\psi(1))\\
&=c_{1}+c_{2}-c_{1}c_{2}.
\end{split}
\end{equation}
This is because the system must be one of the Markov states, i.e., $\pi_{0,0}+ \pi_{1,0}+\phi(1)+ \pi_{0,1}+\psi(1)=1$. $\bar R_{2}$ is consistent with \eqref{equ_etg}.

\section{Proof of Theorem 2}
\label{app_2}
Let $\theta=\sum_i\tau_{i,i}$. The steady-state probabilities of states $[\emptyset,\emptyset]$, $[\mathbf 1_{1},\emptyset]$, $[\emptyset,2\times \mathbf 1_{1}]$ and $[\mathbf 1_{1},2\times \mathbf 1_{1}]$, denoted by $\tau_{0,0}$, $\tau_{1,0}$, $\tau_{0,1}\ \text{and}\ \tau_{1,1}$, respectively, can be given by
\begin{subequations}\label{equ_steady1}\begin{align}
\tau_{0,0}&=(1-\theta+\tau_{0,0})(1-c_{1})(1-c_{2});\label{equ_steady1_1}\\
\tau_{1,0}&=(1-\theta+\tau_{0,0})c_{1}(1-c_{2});\label{equ_steady1_2}\\
\tau_{0,1}&=(1-\theta+\tau_{0,0})(1-c_{1})c_{2};\label{equ_steady1_3}\\
\tau_{1,1}&=(1-\theta+\tau_{0,0})c_{1}c_{2}+(1-c_{1})(1-c_{2})\tau_{1,1},\label{equ_steady1_4}
\end{align}
\end{subequations}
where $(1-\theta)=\sum_{i=0}^{\infty}(\tau_{i,i+1}+\tau_{i+1,i})$ and the rest of the steps are self-explanatory.

The steady-state probabilities of states $[\mathbf 1_{i+1},2\times \mathbf 1_{i+1}]$, $[\mathbf 1_{i+1},2\times \mathbf 1_{i}]$, and $[\mathbf 1_{i},2\times \mathbf 1_{i+1}]$ ($i\ge1$), denoted by $\tau_{i+1,i+1}$, $\tau_{i+1,i}$, and $\tau_{i,i+1}$, respectively, can be written in recurrence expressions, as given  by
\begin{equation}\label{equ_piii}\begin{split}
\tau_{i+1,i+1}
&=c_{1}c_{2}\tau_{i,i}+(1-c_{1})(1-c_{2})\tau_{i+1,i+1}\\
&=\frac{c_{1}c_{2}}{1-(1-c_{1})(1-c_{2})}\tau_{i,i};\\
\tau_{i+1,i}&=c_{1}(1-c_{2})\tau_{i,i};\\
\tau_{i,i+1}&=(1-c_{1})c_{2}\tau_{i,i},\text{ for }i\ge 1.
\end{split}\end{equation}
As a result, $\theta$ can be obtained by
\begin{equation}\begin{split}\label{equ_steady2}
\theta&=\tau_{0,0}+\frac{1-(1-c_{1})(1-c_{2})}{1-(1-c_{1})(1-c_{2})-c_{1}c_{2}}\tau_{1,1}.
\end{split}\end{equation}

Despite there are an infinite number of states in the FDTMC model (as in Section \ref{sec_genmodel}), we are able to recursively calculate and solve for closed-form expressions. By jointly solving  \eqref{equ_steady1} and \eqref{equ_steady2}, we have a simple and closed-form expression for $\tau_{0,0}$, as given by
\begin{equation}\begin{split}
\label{equ_pimin4}
\tau_{0,0}&=\frac{(c_{1}+c_{2}-2c_{1}c_{2})(1-c_{1})(1-c_{2})}{c_{1}+c_{2}-c_{1}c_{2}},
\end{split}
\end{equation}
which essentially provides the probability of strong consistency in a public blockchain, before synchronization is conducted between the miners.

For the rest of the states $[\mathbf 1_{i},2\times \mathbf 1_{i}]$, $[\mathbf 1_{i},2\times \mathbf 1_{i-1}]$, and $[\mathbf 1_{i-1},2\times \mathbf 1_{i}]$ $(i\ge1)$, their steady-state probabilities can be obtained by jointly solving \eqref{equ_steady1}, \eqref{equ_piii}, \eqref{equ_steady2} and \eqref{equ_pimin4}:
\begin{equation}
\label{equ_steady3}
\begin{split}
&\tau_{i,i}=\frac{(c_{1}c_{2})^i(c_{1}+c_{2}-2c_{1}c_{2})}{(c_{1}+c_{2}-c_{1}c_{2})^{i+1}};\\
&\tau_{i,i-1}=\frac{c_{1}^{i}c_{2}^{i-1}(1-c_{2})(c_{1}+c_{2}-2c_{1}c_{2})}{(c_{1}+c_{2}-c_{1}c_{2})^{i}};\\
&\tau_{i-1,i}=\frac{c_{1}^{i-1}c_{2}^i(1-c_{1})(c_{1}+c_{2}-2c_{1}c_{2})}{(c_{1}+c_{2}-c_{1}c_{2})^{i}}.
\end{split}
\end{equation}

We note that $\tau_{0,0}+\sum_{i=1}(\tau_{i,i-1}+\tau_{i-1,i})$ gives the strong consistency probability with the effective synchronization taken into account. This is because the miners in states $[\mathbf 1_{i},2\times \mathbf 1_{i-1}]$, and $[\mathbf 1_{i-1},2\times \mathbf 1_{i}]$ have local chains with different lengths, and they can reach strong consistency after synchronization, i.e.,  transit to $[\emptyset,\emptyset]$. The probability of strong consistency of the entire blockchain system after synchronization is therefore given by
\begin{equation}\label{equ_eta}\begin{split}
\eta&=\tau_{0,0}+\sum_{i=1}(\tau_{i,i-1}+\tau_{i-1,i})=\frac{c_{1}+c_{2}-2c_{1}c_{2}}{c_{1}+c_{2}-c_{1}c_{2}}.
\end{split}
\end{equation}

\bibliography{double}

\begin{thebibliography}{10}
\expandafter\ifx\csname url\endcsname\relax
  \def\url#1{\texttt{#1}}\fi
\expandafter\ifx\csname urlprefix\endcsname\relax\def\urlprefix{URL }\fi
\expandafter\ifx\csname href\endcsname\relax
  \def\href#1#2{#2} \def\path#1{#1}\fi

\bibitem{WANG201910}
X.~Wang, X.~Zha, W.~Ni, R.~P. Liu, Y.~J. Guo, X.~Niu, K.~Zheng, Survey on
  blockchain for internet of things, Computer Communications 136 (2019) 10 --
  29.

\bibitem{Nakamoto:2008ti}
S.~Nakamoto, \href{https://bitcoin.org/bitcoin.pdf}{{Bitcoin: A peer-to-peer
  electronic cash system}} (2008).
\newline\urlprefix\url{https://bitcoin.org/bitcoin.pdf}

\bibitem{moro2020distributed}
E.~P. Moro, A.~K. Duke, Distributed ledger technologies and the internet of
  things: A devices attestation system for smart cities, The Journal of The
  British Blockchain Association (2020) 12500.

\bibitem{kogias2016enhancing}
E.~K. Kogias, P.~Jovanovic, N.~Gailly, I.~Khoffi, L.~Gasser, B.~Ford, Enhancing
  {Bitcoin} security and performance with strong consistency via collective
  signing, in: Proc. USENIX Security Symp. (USENIX Security~'16), Austin, TX,
  USA, 2016, pp. 279--296.

\bibitem{wood2014ethereum}
G.~Wood,
  \href{http://www.cryptopapers.net/papers/ethereum-yellowpaper.pdf}{Ethereum:
  A secure decentralised generalised transaction ledger} (Apr. 2014).
\newline\urlprefix\url{http://www.cryptopapers.net/papers/ethereum-yellowpaper.pdf}

\bibitem{10.1007/978-3-662-46803-6_10}
J.~Garay, et~al., {The Bitcoin backbone protocol: Analysis and applications},
  in: Proc. of the 34th Annu. Int. Conf. on the Theory and Appl. of
  Cryptographic Techn. (EUROCRYPT 2015), 2015, pp. 281--310.

\bibitem{king2012ppcoin}
S.~King, S.~Nadal,
  \href{https://peercoin.net/assets/paper/peercoin-paper.pdf}{Ppcoin:
  Peer-to-peer crypto-currency with proof-of-stake} (Aug. 2012).
\newline\urlprefix\url{https://peercoin.net/assets/paper/peercoin-paper.pdf}

\bibitem{Bentov:2014:PAE:2695533.2695545}
I.~Bentov, C.~Lee, A.~Mizrahi, M.~Rosenfeld, {Proof of Activity: Extending
  Bitcoin's Proof of Work via Proof of Stake}, SIGMETRICS Perform. Eval. Rev.
  42~(3) (2014) 34--37.
\newblock \href {http://dx.doi.org/10.1145/2695533.2695545}
  {\path{doi:10.1145/2695533.2695545}}.

\bibitem{Castro:1999te}
M.~Castro, B.~Liskov, {Practical Byzantine} fault tolerance, in: Proc. 3rd
  USENIX Symp. Operating Syst. Design and Implementation (OSDI~'99), Berkeley,
  CA, USA, 1999, pp. 173--186.

\bibitem{vogels2009eventually}
W.~Vogels, Eventually consistent, Commun. ACM 52~(1) (2009) 40--44.

\bibitem{6688704}
C.~Decker, R.~Wattenhofer, {Information propagation in the Bitcoin network},
  in: Proc. 13rd IEEE Int. Conf. on Peer-to-Peer Comput. (P2P' 13), 2013, pp.
  1--10.

\bibitem{7931680}
X.~Zha, et~al., Collaborative authentication in decentralized dense mobile
  networks with key predistribution 12~(10) (2017) 2261--2275.
\newblock \href {http://dx.doi.org/10.1109/TIFS.2017.2705584}
  {\path{doi:10.1109/TIFS.2017.2705584}}.

\bibitem{7423672}
F.~Tschorsch, B.~Scheuermann, Bitcoin and beyond: A technical survey on
  decentralized digital currencies, IEEE Commun. Surveys Tuts. 18~(3) (2016)
  2084--2123.
\newblock \href {http://dx.doi.org/10.1109/COMST.2016.2535718}
  {\path{doi:10.1109/COMST.2016.2535718}}.

\bibitem{piriou2018simulation}
P.-Y. Piriou, J.-F. Dumas, Simulation of stochastic blockchain models, in: 2018
  14th European Dependable Computing Conference (EDCC), IEEE, 2018, pp.
  150--157.

\bibitem{YANG2020101956}
R.~Yang, X.~Chang, J.~Mišić, V.~B. Mišić,
  \href{https://www.sciencedirect.com/science/article/pii/S0167404820302327}{Assessing
  blockchain selfish mining in an imperfect network: Honest and selfish miner
  views}, Computers \& Security 97 (2020) 101956.
\newblock \href {http://dx.doi.org/https://doi.org/10.1016/j.cose.2020.101956}
  {\path{doi:https://doi.org/10.1016/j.cose.2020.101956}}.
\newline\urlprefix\url{https://www.sciencedirect.com/science/article/pii/S0167404820302327}

\bibitem{YU2020101934}
G.~Yu, X.~Zha, X.~Wang, W.~Ni, K.~Yu, J.~A. Zhang, R.~P. Liu, A unified
  analytical model for proof-of-x schemes, Computers \& Security 96 (2020)
  101934.

\bibitem{Gervais:2016tu}
A.~Gervais, et~al., On the security and performance of proof of work
  blockchains, in: Proc. 23rd ACM SIGSAC Conf. on Comput. and Commun. Security.
  (CCS~'16), ACM, 2016, pp. 3--16.
\newblock \href {http://dx.doi.org/10.1145/2976749.2978341}
  {\path{doi:10.1145/2976749.2978341}}.

\bibitem{sompolinsky2015secure}
Y.~Sompolinsky, A.~Zohar, Secure high-rate transaction processing in {Bitcoin},
  in: Proc. 19th Int. Conf. Financial Cryptogr. Data Secur. (FC~'15), Malta,
  2015, pp. 507--527.

\bibitem{8946275}
Y.~Shahsavari, K.~Zhang, C.~Talhi, A theoretical model for fork analysis in the
  bitcoin network, in: 2019 IEEE International Conference on Blockchain
  (Blockchain), 2019, pp. 237--244.
\newblock \href {http://dx.doi.org/10.1109/Blockchain.2019.00038}
  {\path{doi:10.1109/Blockchain.2019.00038}}.

\bibitem{10.1145/3412341}
N.~Ruan, D.~Zhou, W.~Jia, \href{https://doi.org/10.1145/3412341}{Ursa: Robust
  performance for <i>nakamoto</i> consensus with self-adaptive throughput}, ACM
  Trans. Internet Technol. 20~(4).
\newblock \href {http://dx.doi.org/10.1145/3412341}
  {\path{doi:10.1145/3412341}}.
\newline\urlprefix\url{https://doi.org/10.1145/3412341}

\bibitem{Gilbert:2002il}
S.~Gilbert, N.~Lynch, {Brewer's conjecture and the feasibility of consistent,
  available, partition-tolerant web services}, ACM SIGACT News 33~(2) (2002)
  51--59.

\bibitem{6127847}
D.~Abadi, Consistency tradeoffs in modern distributed database system design:
  {CAP} is only part of the story 45~(2) (2012) 37--42.
\newblock \href {http://dx.doi.org/10.1109/MC.2012.33}
  {\path{doi:10.1109/MC.2012.33}}.

\bibitem{rahman2017characterizing}
M.~R. Rahman, et~al., Characterizing and adapting the consistency-latency
  tradeoff in distributed key-value stores, ACM Trans. Autonomous \& Adaptive
  Sys. 11~(4) (2017) 20.

\bibitem{6133253}
E.~Brewer, {CAP twelve years later: How the `rules' have changed} 45~(2) (2012)
  23--29.
\newblock \href {http://dx.doi.org/10.1109/MC.2012.37}
  {\path{doi:10.1109/MC.2012.37}}.

\bibitem{kleppmann2015critique}
A.~D. Fekete, K.~Ramamritham, Consistency Models for Replicated Data, Springer,
  Berlin, Heidelberg, 2010, pp. 1--17.

\bibitem{sirer2016bitcoin}
E.~Sirer,
  \href{https://hackingdistributed.com/2016/03/01/bitcoin-guarantees-strong-not-eventual-consistency/}{Bitcoin
  guarantees strong, not eventual, consistency}, Hacking, Distributed.
\newline\urlprefix\url{https://hackingdistributed.com/2016/03/01/bitcoin-guarantees-strong-not-eventual-consistency/}

\bibitem{wang2016virus}
X.~Wang, et~al., Virus propagation modeling and convergence analysis in
  large-scale networks 11~(10) (2016) 2241--2254.

\bibitem{10.1145/3308897.3308956}
M.~Alharby, A.~van Moorsel,
  \href{https://doi.org/10.1145/3308897.3308956}{Blocksim: A simulation
  framework for blockchain systems}, SIGMETRICS Perform. Eval. Rev. 46~(3)
  (2019) 135–138.
\newblock \href {http://dx.doi.org/10.1145/3308897.3308956}
  {\path{doi:10.1145/3308897.3308956}}.
\newline\urlprefix\url{https://doi.org/10.1145/3308897.3308956}

\bibitem{8845253}
Y.~{Aoki}, K.~{Otsuki}, T.~{Kaneko}, R.~{Banno}, K.~{Shudo}, Simblock: A
  blockchain network simulator, in: IEEE INFOCOM 2019 - IEEE Conference on
  Computer Communications Workshops (INFOCOM WKSHPS), 2019, pp. 325--329.
\newblock \href {http://dx.doi.org/10.1109/INFCOMW.2019.8845253}
  {\path{doi:10.1109/INFCOMW.2019.8845253}}.

\bibitem{li2019markov}
Q.-L. Li, J.-Y. Ma, Y.-X. Chang, F.-Q. Ma, H.-B. Yu, Markov processes in
  blockchain systems, Computational Social Networks 6~(1) (2019) 1--28.

\bibitem{motlagh2020modeling}
S.~G. Motlagh, J.~Misic, V.~B. Misic, Modeling of churn process in bitcoin
  network, in: 2020 International Conference on Computing, Networking and
  Communications (ICNC), IEEE, 2020, pp. 686--691.

\bibitem{WANG2019100109}
X.~Wang, G.~Yu, X.~Zha, W.~Ni, R.~P. Liu, Y.~J. Guo, K.~Zheng, X.~Niu, Capacity
  of blockchain based internet-of-things: Testbed and analysis, Internet of
  Things 8 (2019) 100109.

\bibitem{103043}
D.~L. Mills, Internet time synchronization: the network time protocol, {IEEE}
  Trans. Commun. 39~(10) (1991) 1482--1493.
\newblock \href {http://dx.doi.org/10.1109/26.103043}
  {\path{doi:10.1109/26.103043}}.

\bibitem{durrett2010probability}
R.~Durrett, Probability: theory and examples, Cambridge university press, 2010.

\bibitem{konstantopoulos2009markov}
T.~Konstantopoulos,
  \href{https://pdfs.semanticscholar.org/251b/3a804be62da4e0835fcffa704cacb03ce310.pdf}{Markov
  chains and random walks} (2009).
\newline\urlprefix\url{https://pdfs.semanticscholar.org/251b/3a804be62da4e0835fcffa704cacb03ce310.pdf}

\bibitem{kleinrock1976queueing}
L.~Kleinrock, Queueing systems, volume 2: Computer Appl., Vol.~66, wiley, New
  York, 1976.

\bibitem{kreinin1997queueing}
A.~Y. Kreinin, Queueing systems with renovation, Int. J. Stochastic Analysis
  10~(4) (1997) 431--441.

\bibitem{TOWSLEY1991353}
D.~Towsley, S.~K. Tripathi, A single server priority queue with server failures
  and queue flushing, Operations Research Lett. 10~(6) (1991) 353 -- 362.
\newblock \href
  {http://dx.doi.org/https://doi.org/10.1016/0167-6377(91)90008-D}
  {\path{doi:https://doi.org/10.1016/0167-6377(91)90008-D}}.

\bibitem{Indyk:1998:ANN:276698.276876}
P.~Indyk, R.~Motwani, Approximate nearest neighbors: towards removing the curse
  of dimensionality, in: Proc. of the 30th annual ACM Symp. on Theory of
  Comput., ACM, 1998, pp. 604--613.

\bibitem{buterin2017casper}
V.~Buterin, V.~Griffith, Casper the friendly finality gadget, arXiv preprint
  arXiv:1710.09437.

\bibitem{sompolinsky2016spectre}
Y.~Sompolinsky, Y.~Lewenberg, A.~Zohar, Spectre: A fast and scalable
  cryptocurrency protocol., IACR Cryptol. ePrint Arch. 2016 (2016) 1159.

\end{thebibliography}

\end{document}